\documentclass[aps,prb,twocolumn,groupedaddress]{revtex4-2}

\usepackage[dvipdfmx]{graphicx}
\usepackage{amsmath,amssymb,bm,amsthm,amsfonts}

\begin{document}

\title{Magnetic ground states of highly doped two-leg Hubbard ladders with a particle bath}

\author{Hiroaki Onishi}
\email{onishi.hiroaki@jaea.go.jp}
\affiliation{%
Advanced Science Research Center,
Japan Atomic Energy Agency,
Tokai, Ibaraki 319-1195, Japan
}
\author{Seiji Miyashita}
\affiliation{%
JSR-UTokyo Collaboration Hub, CURIE,
Department of Physics, Graduate School of Science,
The University of Tokyo,
7-3-1 Hongo, Bunkyo-Ku, Tokyo 113-0033, Japan
}

\date{\today}

\begin{abstract}
We investigate the ground-state magnetism of a Hubbard model
in a system consisting of a main frame (subsystem) and a particle bath (center sites).
The hole doping in the main frame is controlled by adjusting the chemical potential of the particle bath.
In the weakly doped region,
the saturated ferromagnetic state emerges due to the Nagaoka mechanism [Phys. Rev. B \textbf{90}, 224426 (2014)].
However,
in the highly doped region,
a variety of intriguing magnetic states are observed,
including partially polarized states and nonmagnetic states.
To understand these states,
we analyze the state of the subsystem
by comparing its properties with those of a two-leg ladder system,
which corresponds to the subsystem with the center sites removed.
Furthermore,
to gain insight into the microscopic origin of the magnetic phase diagram,
we study the ground state of the corresponding effective $t$-$J$ model,
derived from the Hubbard model by considering the second-order processes of the electron hopping.
The phase diagram is well reproduced by the effective $t$-$J$ model,
which includes the three-site pair-hopping term.
To elucidate the competition among different energy contributions,
such as potential, kinetic, and magnetic energies,
we classify the energies within the effective $t$-$J$ model.
\end{abstract}

\maketitle

\section{INTRODUCTION}

The ferromagnetism of itinerant electron systems has attracted continuing interest
from both fundamental and application viewpoints.
The Hubbard model was introduced to study the ferromagnetism
in transition metals such as Fe, Co, and Ni
\cite{Hubbard1963,Kanamori1963,Gutzwiller1963},
and since then it has been widely used as the simplest model that captures essential properties of correlated electrons.
The Stoner criterion, within mean-field theory, shows that
the ferromagnetism occurs if the Coulomb interaction and/or the density of states at the Fermi level
are large enough
\cite{Stoner1938}.
However, it overestimates the stability of the ferromagnetism due to the mean-field treatment of electron correlations.
Thus,
it is highly significant to clarify in what conditions the ferromagnetism occurs
by nonperturbative and unbiased approaches.
Since the large Coulomb interaction is a key factor for the ferromagnetism,
the extreme case of infinitely large Coulomb interaction is a useful starting point
for exploring the ferromagnetism.
The Nagaoka ferromagnetism is a vital example
\cite{Nagaoka1966,Thouless1965,Tasaki1989},
where
the saturated ferromagnetic (FM) ground state is rigorously proven to occur
in the condition that
the Coulomb interaction is infinite
and there is one hole added to the half-filling
in appropriate lattices satisfying the connectivity.

Whether the Nagaoka ferromagnetism survives
in the case with finite holes and finite Coulomb interaction
has been examined extensively
\cite{Takahashi1982,Doucot1989,Riera1989,Fang1989,Shastry1990,Basile1990,Doucot1990,Barbieri1990,Toth1991,Suto1991,Linden1991,Hanisch1993,Hanisch1995,Wurth1996,Hanisch1997,Watanabe1997a,Watanabe1997b,Watanabe1999,Sakamoto1996,Daul1997,Daul1998,Liang1995,Kohno1997,Becca2001,Carleo2011,Liu2012,Yun2021,Yun2023}.
With two holes,
the saturated FM state is not the ground state in periodic boundary conditions
\cite{Doucot1989,Fang1989},
while it turns to the ground state in mixed boundary conditions
\cite{Riera1989},
indicating the difficulty in obtaining conclusive results
about the true ground state in the thermodynamic limit.
It has been widely accepted that
the saturated FM state becomes unstable
when the hole concentration exceeds a critical hole density
or the Coulomb interaction decreases below a critical strength.
Variational wavefunction approaches have been developed to evaluate
the critical values of the hole density and the Coulomb interaction
and map out where the FM region extends in the ground-state phase diagram
\cite{Shastry1990,Basile1990,Linden1991,Hanisch1993,Hanisch1995,Wurth1996,Hanisch1997}.
There have also been a number of numerical studies using
exact diagonalization
\cite{Watanabe1997a,Watanabe1997b,Watanabe1999},
density-matrix renormalization group (DMRG)
\cite{Sakamoto1996,Daul1997,Daul1998,Liang1995,Kohno1997,Liu2012},
quantum Monte Carlo (QMC)
\cite{Becca2001,Carleo2011,Yun2021,Yun2023}, etc.
It has been demonstrated that,
for a square lattice with one hole,
the saturated FM state occurs
when the Coulomb interaction is above a critical value in finite lattices,
and the critical Coulomb interaction diverges with the system size linearly
\cite{Yun2021}.
In the thermodynamic limit,
the critical Coulomb interaction diverges
in inverse proportion to the hole density
as the hole density goes to zero
\cite{Linden1991}.

\begin{figure}[t] 
\centering
\includegraphics[clip,scale=0.65]{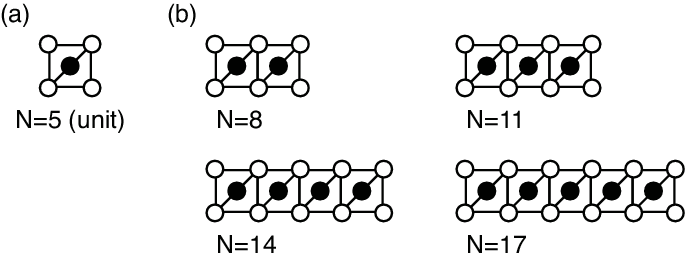}
\caption{
(a) Lattice unit structure.
Open circles represent a subsystem,
and a solid circle at the center represents a site which works as a particle bath.
Solid lines denote the bond for the electron hopping between sites.
For the bond between the subsystem and the center site,
we include only diagonally right up bonds to avoid complexity due to geometrical frustration.
(b) Extended lattices constructed by arranging unit structures in one direction.
}
\label{Fig_lattice}
\end{figure}

On the other hand,
the half-filled system shows a Mott state,
where antiferromagnetic (AFM) correlations occur
due to an effective AFM exchange interaction,
while the total spin is zero in bipartite lattices with the same site number in each sublattice
\cite{Lieb-Mattis1962}.
Thus, the ground state drastically changes
from Mott AFM with the zero total spin to Nagaoka FM with the maximum total spin by adding one hole.
As mentioned,
there have been many works on the cases with more than one hole.
The effect of hole doping has been examined
by changing the number of holes one by one in finite lattices.
In contrast, we have tackled this problem from a viewpoint of gradual hole doping
\cite{Miyashita2008,Onishi2014,Onishi2022,Onishi2023}.
For this purpose,
we have introduced a model
in which we imitate the hole doping.
The system consists of two parts, as shown in Fig.~\ref{Fig_lattice}.
One is a main frame (open circles),
which we call subsystem.
The other part is a particle bath (solid circles),
made up of center sites in plaquettes of the subsystem.
The hole density in the subsystem is controlled continuously
by changing the chemical potential of the center sites.
We regard this situation as a hole doping into the subsystem.
If the subsystem is half-filled and the center sites are empty,
the ground state is the Mott state.
When electrons come to the center sites,
i.e., holes are doped into the subsystem,
a FM ground state appears due to the Nagaoka mechanism
in a finite range of the hole density in the subsystem,
which we call extended Nagaoka ferromagnetism.
We have extensively studied
the change from the Mott state to the FM state in the weakly doped region
\cite{Miyashita2008,Onishi2014,Onishi2022,Onishi2023}.

However,
when the amount of hole doping increases,
qualitatively different phenomena would emerge,
which correspond to the cases with multi-holes.
Indeed, when each center site catches an electron,
the present model is regarded as a kind of Kondo lattice
where the center sites take the role of localized magnetic impurities.
Magnetic properties of such a highly doped system are different
from those of the simple saturated ferromagnetism,
and the system exhibits a rich structure of various phases.
Phenomena in this region have not been considered well.
In our previous work
\cite{Onishi2014},
we left this issue open because of the complicated changes.
There are still no systematic studies and no common understanding about the mechanisms
that produce various magnetic states after the breakdown of the extended Nagaoka FM state.
There, the low-lying energy levels are very close due to the competition among different phases,
which requires careful analyses
compared to the case of the simple saturated ferromagnetism.

In the present paper,
we explore a highly doped region
where the extended Nagaoka FM state disappears due to an excess hole doping over its threshold.
We study the magnetic ground state in this highly doped region
by numerical methods such as Lanczos diagonalization.
To clarify the mechanisms for producing various magnetic states,
we analyze the state of the subsystem
by referring to the property of a two-leg ladder system
that corresponds to the part of the subsystem excluding the center sites.
Here we stress again that in the present model,
we can control the hole density in the subsystem continuously,
by which we grasp the overall nature of the effect of the hole doping
in a wide range of hole doping rate.

Note that,
when the Coulomb interaction is sufficiently large compared with the electron hopping,
the double occupancy is suppressed,
and we can derive an effective $t$-$J$ model
in the restricted Hilbert space without double occupancy.
Considering the second-order processes with respect to the electron hopping,
the effective $t$-$J$ model includes the terms of
the electron hopping,
the exchange interaction among spins,
and the three-site pair-hopping,
as well as the chemical potential of the center sites.
The three-site pair-hopping is often discarded when treating the $t$-$J$ model,
although it has contributions in the same order as the exchange interaction.
The prior researches using the three-site pair-hopping have mainly focused on
the relevance to the superconductivity
\cite{Spalek1988,Ammon1995,Yokoyama1996,Saiga2002,Coulthard2018}.
In this work,
we address
that the term has significant effect on the ferromagnetism.
To elucidate the competition among different energy contributions,
such as potential, kinetic, and magnetic energies,
we classify the energies within the effective $t$-$J$ model.

The organization of the present paper is as follows.
In Sec.~II,
we describe a model for the extended Nagaoka ferromagnetism,
and also explain numerical methods.
In Sec.~III,
we investigate the magnetic ground state,
mainly focusing on the region
where holes are highly doped into the subsystem.
In Sec.~IV,
we examine the ground state of the effective $t$-$J$ model.
In Sec.~V,
we show results on the dependence on the system size.
Section~VI is devoted to summary and discussion.

\section{MODEL AND NUMERICAL METHOD}

We have introduced the Hubbard model including sites of the particle bath,
by which we imitate the hole doping into the system
via the change of electron distribution controlled by the chemical potential of the particle bath
\cite{Miyashita2008,Onishi2014,Onishi2022,Onishi2023}.
As a unit, we take a five-site system depicted in Fig.~\ref{Fig_lattice}(a).
The system consists of two parts:
A four-site plaquette (open circles) is the subsystem,
while an additional site at the center (solid circle) works as the particle bath.
We put four electrons
to have a half-filled situation of the subsystem
when all electrons sit in the subsystem,
yielding the Mott state.
If an electron escapes to the center site,
the subsystem has one hole and thereby the Nagaoka ferromagnetism occurs.

We construct extended lattices by arranging the five-site units in one direction,
as shown in Fig.~\ref{Fig_lattice}(b).
We use open boundary conditions.
Note that arrangements in two and three dimensions show qualitatively similar results~\cite{Onishi2023}.
The model is described by
\begin{align}
  H
  =&
  -t \sum_{\langle i,j \rangle,\sigma} (c_{i\sigma}^{\dagger}c_{j\sigma}+{\mathrm{H.c.}})
  +U \sum_{i} n_{i\uparrow}n_{i\downarrow}
\nonumber \\
  &
  +\mu \sum_{i \in {\mathrm{center}}} n_{i},
\label{Eq_H}
\end{align}
where $c_{i\sigma}$ is an annihilation operator of an electron with spin $\sigma$ $(=\uparrow,\downarrow)$ at site $i$,
$n_{i\sigma}=c_{i\sigma}^{\dag}c_{i\sigma}$,
$n_{i}=n_{i\uparrow}+n_{i\downarrow}$,
$\langle i,j \rangle$ denotes neighboring sites connected by a bond,
$t$ is the hopping amplitude,
$U$ is the Coulomb interaction at all sites,
and $\mu$ is the on-site potential at the center sites,
which denotes the chemical potential of the particle bath.
Throughout the paper, we set $t=1$ as the energy unit.
The number of sites is $N=N^{\mathrm{sub}}+N^{\mathrm{c}}$,
where $N^{\mathrm{sub}}$ and $N^{\mathrm{c}}$ are the number of sites
in the subsystem and that in the center sites, respectively.
We set the number of electrons $N_{\mathrm{e}}$ equal to $N^{\mathrm{sub}}$
to have a half-filled situation in the subsystem
when all electrons are accommodated in the subsystem.

For large positive $\mu$,
the center sites are empty, and the subsystem is half-filled,
so that the ground state is the Mott state.
As studied in detail in the previous works,
when $\mu$ decreases,
the FM ground state occurs,
which we call extended Nagaoka ferromagnetism
\cite{Miyashita2008,Onishi2014,Onishi2022,Onishi2023}.
As $\mu$ decreases further,
for negative $\mu$,
each center site catches almost one electron as long as $\mu>-U$,
and the number of holes in the subsystem increases up to $N^{\mathrm{c}}$.
In such a case, the electron density in the subsystem is 1/2
in the large $N$ limit.
%
%
Thus, the subsystem is highly doped.
If $\mu<-U$,
the center sites are doubly occupied,
and the subsystem is heavily doped
where the electron density in the subsystem goes down to zero in the large $N$ limit.
In the present paper,
we search for intriguing phenomena in the highly doped region
mainly focusing on the case of $\mu>-U$.
The electron density in the subsystem would be in the range between 1/2 and 1.

In the present work,
we investigate the ground-state property of this region by exploiting numerical methods
such as Lanczos method.
Since the total magnetization,
given by $S_{\mathrm{tot}}^{z}=\sum_{i}S_{i}^{z}$,
is conserved,
we calculate the ground state as the lowest-energy state
in the subspace of $S_{\mathrm{tot}}^{z}=0$.
We study the changes of the ground state by investigating microscopic properties.
We first analyze the system with $N=8$ sites in detail.
After that, we also show numerical results for larger systems with $N=11$, 14, and 17 sites.
In the present Lanczos calculations,
we encountered slow convergence,
which is described in Appendix A.

To characterize the magnetic property of the ground state,
we evaluate the total spin $S_{\mathrm{tot}}$,
given by
\begin{equation}
  S_{\mathrm{tot}}(S_{\mathrm{tot}}+1) = \langle \bm{S}_{\mathrm{tot}}^{2} \rangle,
\label{Eq_Stot2}
\end{equation}
where
$\bm{S}_{\mathrm{tot}}=\sum_{i}\bm{S}_{i}$,
$\bm{S}_{i}=\frac{1}{2}\sum_{\alpha\beta}c_{i\alpha}^{\dag}\bm{\sigma}_{\alpha\beta}c_{i\beta}$,
$\bm{\sigma}$ are Pauli matrices,
and $\langle \cdots \rangle$ denotes the expectation value in the ground state.
From Eq.~(\ref{Eq_Stot2}),
the total spin is calculated as
\begin{equation}
  S_{\mathrm{tot}} = \left( -1+\sqrt{1+4\langle\bm{S}_{\mathrm{tot}}^{2}\rangle} \right)/2.
\label{Eq_Stot}
\end{equation}
We note that $S_{\mathrm{tot}}$ should be an integer,
since there are even numbers of electrons,
i.e., $N_{\mathrm{e}}=N^{\mathrm{sub}}$,
in the present setup.
We also evaluate the total spins of parts of the system, defined by
\begin{equation}
  S_{\mathrm{sub}}(S_{\mathrm{sub}}+1) = \langle\bm{S}_{\mathrm{sub}}^{2}\rangle,
\end{equation}
\begin{equation}
  S_{\mathrm{c}}(S_{\mathrm{c}}+1) = \langle\bm{S}_{\mathrm{c}}^{2}\rangle,
\end{equation}
where
$\bm{S}_{\mathrm{sub}}=\sum_{i\in {\mathrm{subsystem}}}\bm{S}_{i}$
and
$\bm{S}_{\mathrm{c}}=\sum_{i\in {\mathrm{center}}}\bm{S}_{i}$.
We note that $\bm{S}_{\mathrm{sub}}^{2}$ and $\bm{S}_{\mathrm{c}}^{2}$ are not conserved quantities separately,
and $S_{\mathrm{sub}}$ and $S_{\mathrm{c}}$ take general real values.

As for the distribution of electrons,
we measure the number of electrons in the center sites,
\begin{equation}
  N_{\mathrm{e}}^{\mathrm{c}} = \sum_{i \in \mathrm{center}} \langle n_{i} \rangle.
\end{equation}
As mentioned above,
we set the number of electrons $N_{\mathrm{e}}$ equal to the number of sites in the subsystem $N^{\mathrm{sub}}$,
so that $N_{\mathrm{e}}^{\mathrm{c}}$ corresponds to the number of holes doped into the subsystem.

\section{MAGNETIC GROUND STATE}

We investigate the ground state
with a main  focus on the negative $\mu$ region where holes are highly doped into the subsystem.
Although we have reported in detail on the transition
between Mott and extended Nagaoka FM states
in the weakly doped region
\cite{Miyashita2008,Onishi2014,Onishi2022,Onishi2023},
we include such parameter region
for the completeness of the present paper.

\begin{figure}[t] 
\centering
\includegraphics[clip,scale=0.65]{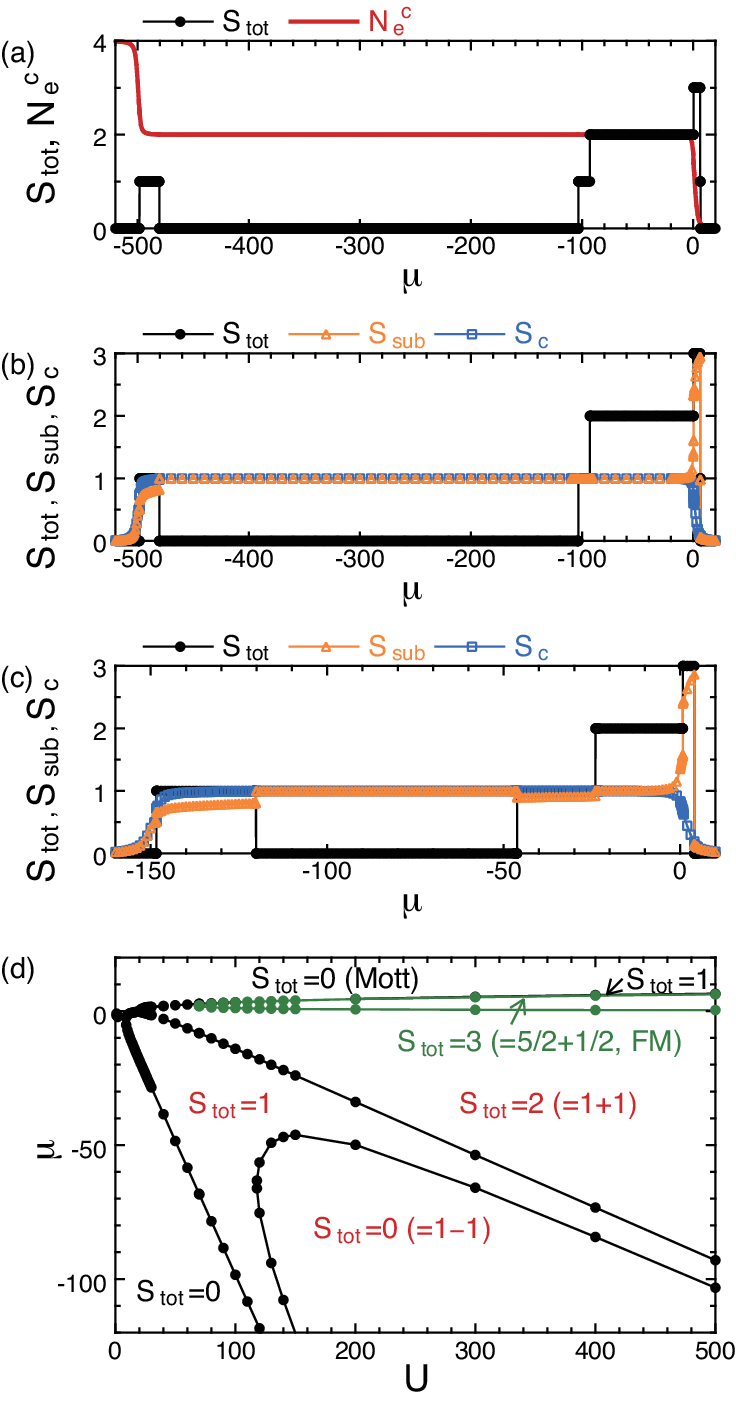}
\caption{
Properties of the eight-site system with six electrons.
(a) The total spin $S_{\mathrm{tot}}$
and the number of electrons in the center sites $N_{\mathrm{e}}^{\mathrm{c}}$
as a function of $\mu$ at $U=500$.
(b) The total spin $S_{\mathrm{tot}}$
together with those in the subsystem $S_{\mathrm{sub}}$ and in the center sites $S_{\mathrm{c}}$
as a function of $\mu$ at $U=500$.
(c) The $\mu$ dependences of $S_{\mathrm{tot}}$, $S_{\mathrm{sub}}$, and $S_{\mathrm{c}}$ at $U=150$.
(d) Ground-state phase diagram.
Solid circles represent points where $S_{\mathrm{tot}}$ changes.
}
\label{Fig_n8}
\end{figure}

In Fig.~\ref{Fig_n8},
we show properties of the eight-site system composed of two units.
In Fig.~\ref{Fig_n8}(a),
we show $S_{\mathrm{tot}}$ and $N_{\mathrm{e}}^{\mathrm{c}}$ as a function of $\mu$ at $U=500$
which is large enough to observe typical $\mu$ dependences in the strong-coupling regime.
The saturated FM state (the extended Nagaoka ferromagnetism) of $S_{\mathrm{tot}}=3$ is realized in $0.4 \leq \mu \leq 6.3$.
In the present range of $\mu$ including large negative values,
the FM state locates in a very narrow range of positive $\mu$
in the right-hand side of the panel of Fig.~\ref{Fig_n8}(a).

Below this region,
$N_{\mathrm{e}}^{\mathrm{c}} \simeq 2$,
i.e., two holes are doped into the subsystems,
and thus we expect that $S_{\mathrm{sub}} \simeq 2$
if four electrons in the subsystem fully polarize.
However,
$S_{\mathrm{sub}} \simeq 1$ and $S_{\mathrm{c}} \simeq 1$,
as shown in Fig.~\ref{Fig_n8}(b).
That is, four electrons in the subsystem partially polarize,
which implies that the Nagaoka FM state in the subsystem is broken due to the excess hole doping.
The remaining two electrons in the center sites polarize,
although the center sites are not connected directly.
This observation suggests that the coupling between the subsystem and the center sites
induces the spin polarization of the center sites.
With further decreasing $\mu$,
$S_{\mathrm{sub}} \simeq 1$ and $S_{\mathrm{c}} \simeq 1$ are unchanged,
while $S_{\mathrm{tot}}$ changes sequentially as 2, 1, and 0,
indicating changes of the way of
the composition of $S_{\mathrm{sub}} \simeq 1$ and $S_{\mathrm{c}} \simeq 1$.

The $S_{\mathrm{tot}}=0$ state where
$N_{\mathrm{e}}^{\mathrm{c}} \simeq 2$, $S_{\mathrm{sub}} \simeq 1$, and $S_{\mathrm{c}} \simeq 1$
keeps stable until $\mu$ decreases down to a value near $-U$.
In a narrow region near $\mu=-U$,
a partial FM state of $S_{\mathrm{tot}}=1$ appears,
while keeping $N_{\mathrm{e}}^{\mathrm{c}} \simeq 2$.
With further decreasing $\mu$,
the ground state finally changes to show the double occupancy in the center sites,
i.e., $N_{\mathrm{e}}^{\mathrm{c}} \simeq 4$,
and the ground state becomes spin singlet
with $S_{\mathrm{sub}} \simeq 0$ and $S_{\mathrm{c}} \simeq 0$.

The value of $U=500$ discussed above is large and well represents the case of $U=\infty$.
When $U$ becomes small,
the magnetic states are affected.
In Fig.~\ref{Fig_n8}(c),
we show the dependences of
$S_{\mathrm{tot}}$, $S_{\mathrm{sub}}$, and $S_{\mathrm{c}}$ at $U=150$.
We find phases of the same set of values of $S_{\mathrm{tot}}$ as that observed at $U=500$
in Fig.~\ref{Fig_n8}(b).
The widths of two regions of $S_{\mathrm{tot}}=1$
above $\mu=-46.2$ and below $\mu=-120.2$
(the edges of the region of $S_{\mathrm{tot}}=0$), increase,
while that of the region of $S_{\mathrm{tot}}=0$ decreases.
Moreover,
in the two regions of $S_{\mathrm{tot}}=1$,
the value of $S_{\mathrm{sub}}$ slightly deviates from one to a non-integer value.
Such non-integer value is allowed for $S_{\mathrm{sub}}$ and $S_{\mathrm{c}}$,
since they are not conserved quantities
in contrast to $S_{\mathrm{tot}}$.
Thus, qualitative features of phases are maintained between $U=500$ and $U=150$,
although we find quantitative changes due to the effects of electron correlations.

In Fig.~\ref{Fig_n8}(d),
we show the ground-state phase diagram.
Below the region of the saturated FM state,
the regions of $S_{\mathrm{tot}}=2$, $S_{\mathrm{tot}}=1$, and $S_{\mathrm{tot}}=0$ widely spread,
where $S_{\mathrm{sub}} \simeq 1$ and $S_{\mathrm{c}} \simeq 1$ are commonly realized,
while $S_{\mathrm{sub}}$ decreases from one when $U$ becomes small.
The region of $S_{\mathrm{tot}}=0$ is surrounded by that of $S_{\mathrm{tot}}=1$,
and its range of $\mu$ shrinks as $U$ decreases,
leading to a reentrant behavior at $U \gtrsim 118$.
The competition among phases in such a region would provide an interesting playground
for quantum dynamics and finite temperatures.
The ground state eventually becomes of $S_{\mathrm{tot}}=0$ below the boundary close to $\mu=-U$.

\begin{figure}[t] 
\centering
\includegraphics[clip,scale=0.65]{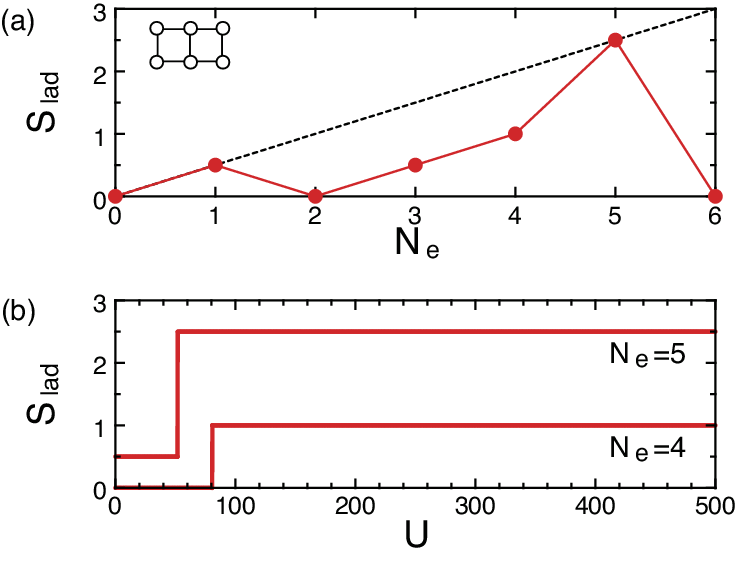}
\caption{
(a) The total spin of the two-leg ladder system $S_{\mathrm{lad}}$
in the limit of infinite $U$
as a function of the number of electrons $N_{\mathrm{e}}$
for $N_{\mathrm{lad}}=6$.
(b) $S_{\mathrm{lad}}$
as a function of $U$
for $N_{\mathrm{lad}}=6$ with $N_{\mathrm{e}}=5$ (one hole) or $N_{\mathrm{e}}=4$ (two holes).
}
\label{Fig_two-leg-ladder_n6}
\end{figure}

In order to have an insight into mechanisms of the various magnetic states in the present model,
we look into what types of magnetic states are possible
in the part of the subsystem, eliminating the center sites.
For this purpose, we investigate a two-leg ladder system.
In Fig.~\ref{Fig_two-leg-ladder_n6},
we show the total spin of the two-leg ladder system $S_{\mathrm{lad}}$
for the case of $N_{\mathrm{lad}}=6$ sites
to compare with the case of $N=8$ sites of the present model.
Figure~\ref{Fig_two-leg-ladder_n6}(a) shows $S_{\mathrm{lad}}$
in the limit of infinite $U$ as a function of the number of electrons.
When there is one hole,
the ground state is the saturated FM state of $S_{\mathrm{lad}}=5/2$
due to the realization of the Nagaoka FM state.
When there are two holes,
a partial FM state of $S_{\mathrm{lad}}=1$ occurs,
while the saturated FM state of $S_{\mathrm{lad}}=2$ is not realized
even in the limit of infinite $U$.
As shown in Fig.~\ref{Fig_two-leg-ladder_n6}(b),
the saturated FM state of $S_{\mathrm{lad}}=5/2$ for $N_{\mathrm{e}}=5$ (one hole)
and
the partial FM state of $S_{\mathrm{lad}}=1$ for $N_{\mathrm{e}}=4$ (two holes)
are realized when $U$ is above a sufficiently large value.
If there are more than two holes,
$S_{\mathrm{lad}}$ takes the minimum value 0 or 1/2
for the cases with even or odd electrons, respectively,
even in the limit of infinite $U$.
That is, the spin polarization does not occur.

Based on these facts,
we consider how the magnetic states of the present model
are formed as the combined states of the subsystem and the center sites.
Regarding the saturated FM state,
the state of the subsystem is dominated by the Nagaoka FM state.
The occurrence of the Nagaoka FM state is indeed confirmed
in the two-leg ladder system with one hole.
At negative $\mu$ where the subsystem has two holes,
$S_{\mathrm{sub}} \simeq 1$ and $S_{\mathrm{c}} \simeq 1$,
causing the states of $S_{\mathrm{tot}}=2$, 1, and 0.
The total spin of the subsystem is rather stable at $S_{\mathrm{sub}} \simeq 1$,
which would be attributed to the fact that
the two-leg ladder system with two holes has $S_{\mathrm{lad}}=1$.
At large negative $\mu$ below $-U$ where the subsystem has four holes,
$S_{\mathrm{sub}} \simeq 0$ and $S_{\mathrm{c}} \simeq 0$ and the ground state is of spin singlet.
The situation that $S_{\mathrm{sub}} \simeq 0$ is consistent with
$S_{\mathrm{lad}}=0$ for $N_{\mathrm{e}}=2$.
From these similarities,
we attribute the characteristics of the state of the subsystem
to those of the ground states realized in the two-leg ladder system with corresponding numbers of holes.

\section{EFFECTIVE $t$-$J$ MODEL}

\begin{figure}[t] 
\centering
\includegraphics[clip,scale=0.55]{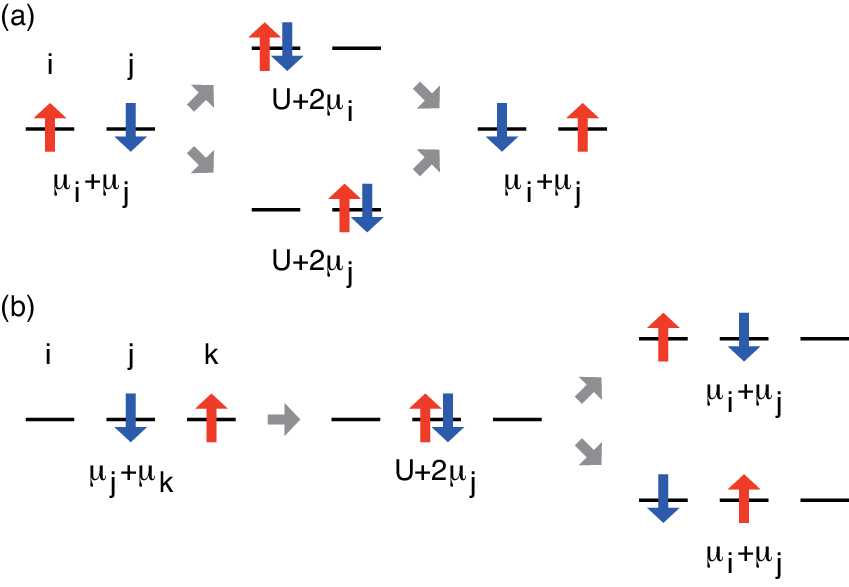}
\caption{
Second-order processes of the electron hopping in the strong-coupling limit.
(a) Exchange and (b) three-site pair-hopping processes.
}
\label{Fig_tJ_process}
\end{figure}

\begin{figure}[t] 
\centering
\includegraphics[clip,scale=0.8]{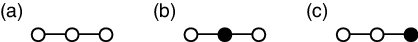}
\caption{
Schematics of neighboring three sites.
Open circles represent the subsystem
and solid circles are the center sites.
(a) Three sites are within the subsystem.
(b) A center site is in the middle of three sites.
(c) A center site is in the side of three sites.
Note that there are no patterns with more than one center site
according to the lattice structure,
as shown in Fig.~\ref{Fig_lattice}.
}
\label{Fig_3-site-config}
\end{figure}

We have shown that magnetic states possessing finite total spin are induced
by holes doped into the subsystem.
In an intuitive picture,
the hole motion under the Pauli exclusion principle would preferably cause parallel spin configurations
to avoid the energy cost of the Coulomb interaction between antiparallel-spin electrons.
However,
a key role of magnetic interactions
in the realization of magnetic states is not clear,
since the present Hubbard model (\ref{Eq_H}) does not include spin-dependent interactions explicitly.
In this context,
it is helpful to construct an effective model
that has interactions of spin degrees of freedom.
Thus, we consider an effective $t$-$J$ model
which is derived by considering the second-order processes of the electron hopping.
Note that
the parameter space is restricted to $U>\vert \mu \vert$
to ensure that the local energy of the singly occupied state is lower than that of the doubly occupied state
for the unperturbed state.
In this constraint,
the Hamiltonian of the effective $t$-$J$ model is given by
\begin{align}
  H_{\mathrm{eff}}
  =&
  -t \sum_{\langle i,j \rangle,\sigma} (\tilde{c}_{i\sigma}^{\dag}\tilde{c}_{j\sigma} + \mathrm{H.c.})
  +\mu \sum_{i \in \mathrm{center}} \tilde{n}_{i}
\nonumber \\
  &
  +\sum_{\langle i,j \rangle}
  J_{ij} \left( \tilde{\bm{S}}_{i} \cdot \tilde{\bm{S}}_{j}-\frac{1}{4}\tilde{n}_{i}\tilde{n}_{j} \right)
\nonumber \\
  &
  -\sum_{\langle\langle i,j,k \rangle\rangle,\sigma}^{i \neq k}
  \frac{J_{ijk}}{4}
  ( \tilde{c}_{i\sigma}^{\dag} \tilde{c}_{j,-\sigma}^{\dag} \tilde{c}_{j,-\sigma} \tilde{c}_{k\sigma}
\nonumber \\
  & \ \ \ \ \ \ \ \ \ \ \ \ \ \ \ \
    -\tilde{c}_{i,-\sigma}^{\dag} \tilde{c}_{j\sigma}^{\dag} \tilde{c}_{j,-\sigma} \tilde{c}_{k\sigma}
  + \mathrm{H.c.}),
\label{Eq_Heff}
\end{align}
where $\tilde{c}_{i\sigma} \equiv c_{i\sigma}(1-n_{i,-\sigma})$ is an annihilation operator of an electron
under the constraint of no double occupancy.
$J_{ij}$ is the exchange interaction
obtained by considering the exchange process in Fig.~\ref{Fig_tJ_process}(a) as
\begin{equation}
  J_{ij}=\frac{4t^2}{U}\frac{1}{1-\{(\mu_{i}-\mu_{j})/U\}^2},
\label{Eq_Jij}
\end{equation}
where $\mu_{i}=\mu$ for the center sites and $\mu_{i}=0$ for the sites in the subsystem.
Concretely, substituting $\mu_{i}$ to Eq.~(\ref{Eq_Jij}),
we obtain
$J_{\mathrm{ex}}^{\mathrm{ss}}=4t^2/U$
for bonds within the subsystem,
and
$J_{\mathrm{ex}}^{\mathrm{sc}}=4t^2/U\{1-(\mu/U)^2\}$
for bonds connecting a site in the subsystem and a center site.
$J_{ijk}$ is the three-site pair-hopping
which is effective when a hole is adjacent to an electron pair with antiparallel spins,
as shown in Fig.~\ref{Fig_tJ_process}(b),
and $\langle\langle i,j,k \rangle\rangle$ stands for neighboring three sites.
The expression of the three-site pair-hopping is given by
\begin{equation}
  J_{ijk}=
  \frac{4t^2}{U}
  \frac{1}{2}
  \left\{
    \frac{1}{1+(\mu_{j}-\mu_{k})/U} + \frac{1}{1+(\mu_{j}-\mu_{i})/U}
  \right\},
\label{Eq_Jijk}
\end{equation}
which leads to
$J_{\mathrm{ph}}^{\mathrm{sss}}=4t^{2}/U$
when three sites are within the subsystem [Fig.~\ref{Fig_3-site-config}(a)],
$J_{\mathrm{ph}}^{\mathrm{scs}}=4t^{2}/U(1+\mu/U)$
when a center site is in the middle of three sites [Fig.~\ref{Fig_3-site-config}(b)],
and
$J_{\mathrm{ph}}^{\mathrm{ssc}}=4t^{2}(1-\mu/2U)/U(1-\mu/U)$
when a center site is in the side of three sites [Fig.~\ref{Fig_3-site-config}(c)].
The three-site pair-hopping is often discarded
in usual treatments of the $t$-$J$ model,
but this term plays an important role to reproduce the property of the original Hubbard model quantitatively,
as we will see below.

\subsection{Simplified $t$-$J$ model}

\begin{figure}[t] 
\centering
\includegraphics[clip,scale=0.65]{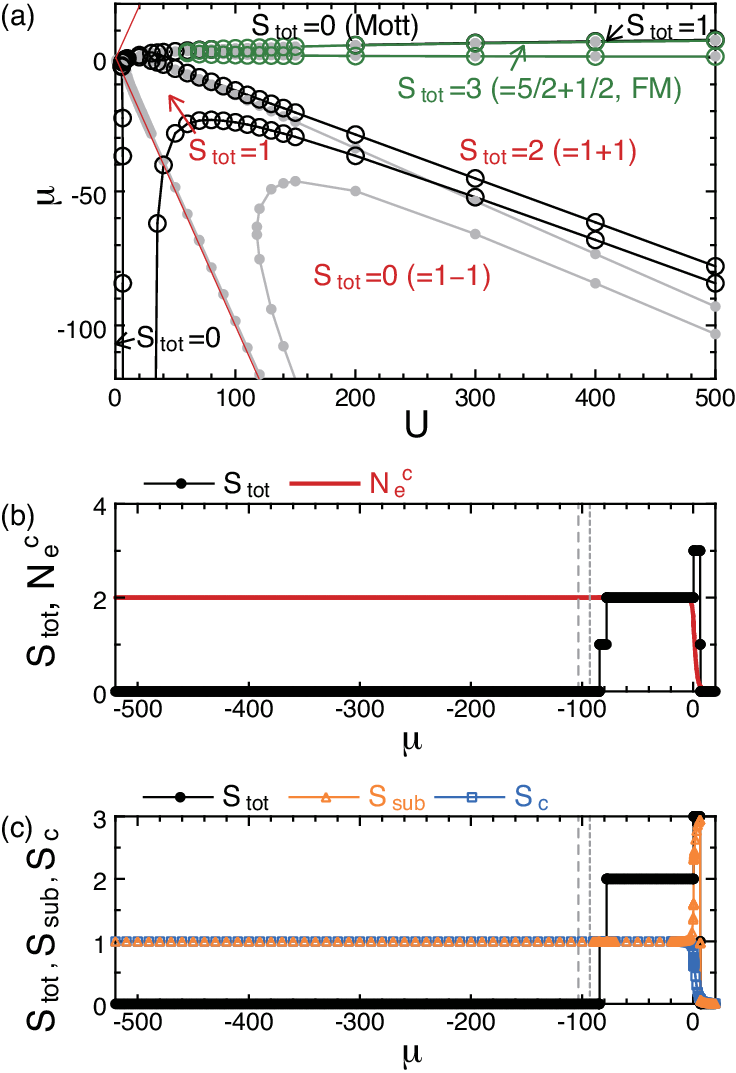}
\caption{
Properties of the simplified $t$-$J$ model (\ref{Eq_Heff_s1}) with $N=8$ and $N_{\mathrm{e}}=6$,
compared to the results of the Hubbard model
presented in Fig.~\ref{Fig_n8}.
(a) Ground-state phase diagram.
Open circles are results of the simplified $t$-$J$ model,
and solid circles in gray are results of the Hubbard model,
given in Fig.~\ref{Fig_n8}(d).
Boundaries of the parameter space $\mu=\pm U$ are denoted by red lines.
(b) The total spin $S_{\mathrm{tot}}$
and the number of electrons in the center sites $N_{\mathrm{e}}^{\mathrm{c}}$
as a function of $\mu$ at $U=500$.
(c) The total spin $S_{\mathrm{tot}}$
together with those in the subsystem $S_{\mathrm{sub}}$ and in the center sites $S_{\mathrm{c}}$
as a function of $\mu$ at $U=500$.
In (b) and (c),
vertical dotted and dashed lines around $\mu \simeq -100$ denote
the transition point from $S_{\mathrm{tot}}=2$ to 1 and
that from $S_{\mathrm{tot}}=1$ to 0, respectively,
in the case of the Hubbard model.
}
\label{Fig_tJ_simple1}
\end{figure}

To grasp the significance of the exchange interaction,
we start with the discussion on a simplified model only with uniform exchange interaction $J$,
described by
\begin{align}
  H_{\mathrm{eff}}^{\prime}
  =&
  -t \sum_{\langle i,j \rangle,\sigma} (\tilde{c}_{i\sigma}^{\dag}\tilde{c}_{j\sigma} + \mathrm{H.c.})
  +\mu \sum_{i \in \mathrm{center}} \tilde{n}_{i}
\nonumber \\
  &
  +J \sum_{\langle i,j \rangle}
  \left( \tilde{\bm{S}}_{i} \cdot \tilde{\bm{S}}_{j}-\frac{1}{4}\tilde{n}_{i}\tilde{n}_{j} \right),
\label{Eq_Heff_s1}
\end{align}
where $J=4t^{2}/U$.
In this simplified model,
we assume $J_{ij}=4t^2/U$ for all bonds,
i.e., we ignore the dependence of $J_{ij}$ on $\mu$,
and $J_{ijk}=0$.
In Fig.~\ref{Fig_tJ_simple1},
we compare the results of the present $t$-$J$ model and those of the original Hubbard model
in the system of $N=8$ and $N_{\mathrm{e}}=6$.
We note that
we can formally analyze the simplified $t$-$J$ model (\ref{Eq_Heff_s1}) for $U>0$,
although the parameter space of $U$ is originally restricted to $U>\vert \mu \vert$
in the effective $t$-$J$ model (\ref{Eq_Heff}).
In this subsection,
we simply show numerical results for the whole range of $U>0$.

In Fig.~\ref{Fig_tJ_simple1}(a),
we show the ground-state phase diagram
of this simplified $t$-$J$ model (\ref{Eq_Heff_s1}).
We find phases of the same set of values of $S_{\mathrm{tot}}$ as that of the Hubbard model (\ref{Eq_H}),
but the difference of phase boundaries is remarkable as $\mu$ decreases to large negative values.
In particular,
a phase boundary
between the regions of $S_{\mathrm{tot}}=1$ and $S_{\mathrm{tot}}=0$
appears near the vertical axis,
in contrast to the phase boundary close to $\mu=-U$
in the case of the Hubbard model.

In Fig.~\ref{Fig_tJ_simple1}(b),
we show the dependences of $S_{\mathrm{tot}}$ and $N_{\mathrm{e}}^{\mathrm{c}}$ on $\mu$ at $U=500$,
i.e., the same condition as that in Fig.~\ref{Fig_n8}(a).
An obvious difference from the Hubbard model is that
$N_{\mathrm{e}}^{\mathrm{c}} \simeq 2$ keeps unchanged
even when $\mu$ decreases down to the values below $-U$.
This is because the double occupancy is prohibited in the $t$-$J$ model,
while $N_{\mathrm{e}}^{\mathrm{c}} \simeq 4$
due to the double occupancy in the center sites
in the Hubbard model.
Moreover,
we do not find a partial FM state of $S_{\mathrm{tot}}=1$
near $\mu=-U$.
On the other hand,
when $\mu$ is sufficiently above $-U$,
the dependences of $S_{\mathrm{tot}}$ and $N_{\mathrm{e}}^{\mathrm{c}}$ are similar to those in the Hubbard model.
The values of $S_{\mathrm{tot}}$ agree with those in the Hubbard model,
whereas transition points around $\mu \simeq -80$ shift to the positive $\mu$ side
from those in the Hubbard model around $\mu \simeq -100$ (vertical dotted and dashed lines).
As shown in Fig.~\ref{Fig_tJ_simple1}(c),
the dependences of $S_{\mathrm{sub}}$ and $S_{\mathrm{c}}$ also resemble those in the Hubbard model,
as compared with Fig.~\ref{Fig_n8}(b).

\subsection{Effect of chemical potential}

\begin{figure}[t] 
\centering
\includegraphics[clip,scale=0.65]{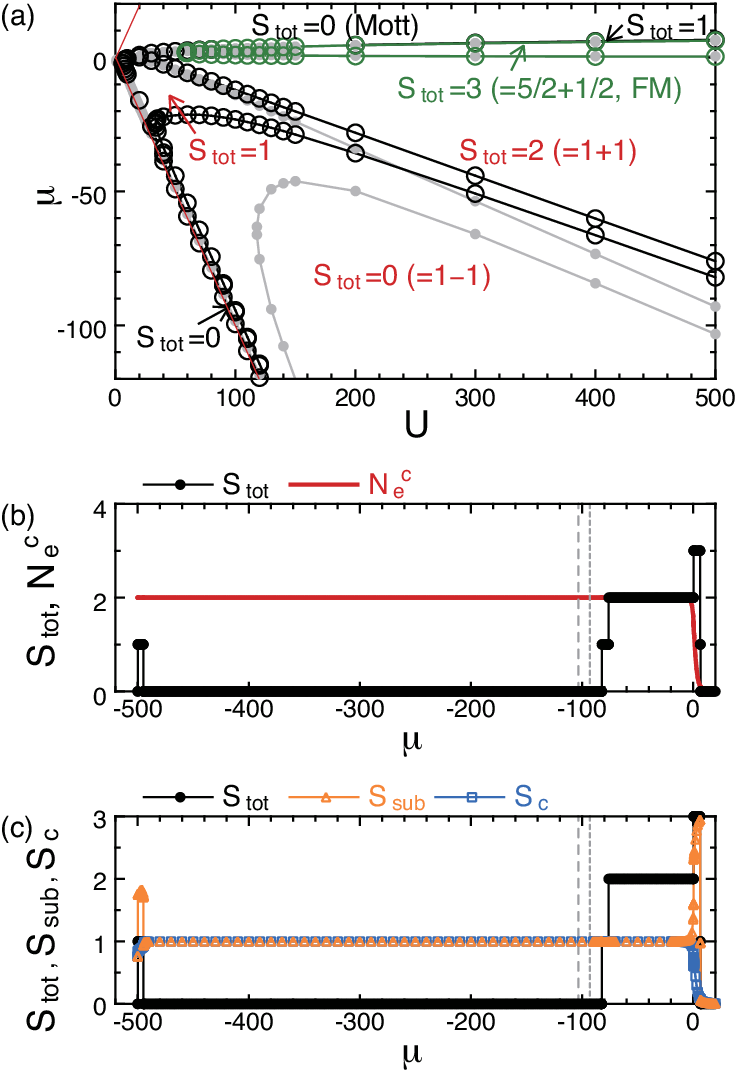}
\caption{
Properties of the simplified $t$-$J$ model (\ref{Eq_Heff_s2}) with $N=8$ and $N_{\mathrm{e}}=6$,
compared to the results of the Hubbard model
presented in Fig.~\ref{Fig_n8}.
(a) Ground-state phase diagram.
Open circles are results of the simplified $t$-$J$ model,
and solid circles in gray are results of the Hubbard model,
given in Fig.~\ref{Fig_n8}(d).
Boundaries of the parameter space $\mu=\pm U$ are denoted by red lines.
(b) The total spin $S_{\mathrm{tot}}$
and the number of electrons in the center sites $N_{\mathrm{e}}^{\mathrm{c}}$
as a function of $\mu$ at $U=500$.
(c) The total spin $S_{\mathrm{tot}}$
together with those in the subsystem $S_{\mathrm{sub}}$ and in the center sites $S_{\mathrm{c}}$
as a function of $\mu$ at $U=500$.
In (b) and (c),
vertical dotted and dashed lines around $\mu \simeq -100$ denote
the transition point from $S_{\mathrm{tot}}=2$ to 1 and
that from $S_{\mathrm{tot}}=1$ to 0, respectively,
in the case of the Hubbard model.
}
\label{Fig_tJ_simple2}
\end{figure}

\begin{figure}[t] 
\centering
\includegraphics[clip,scale=0.65]{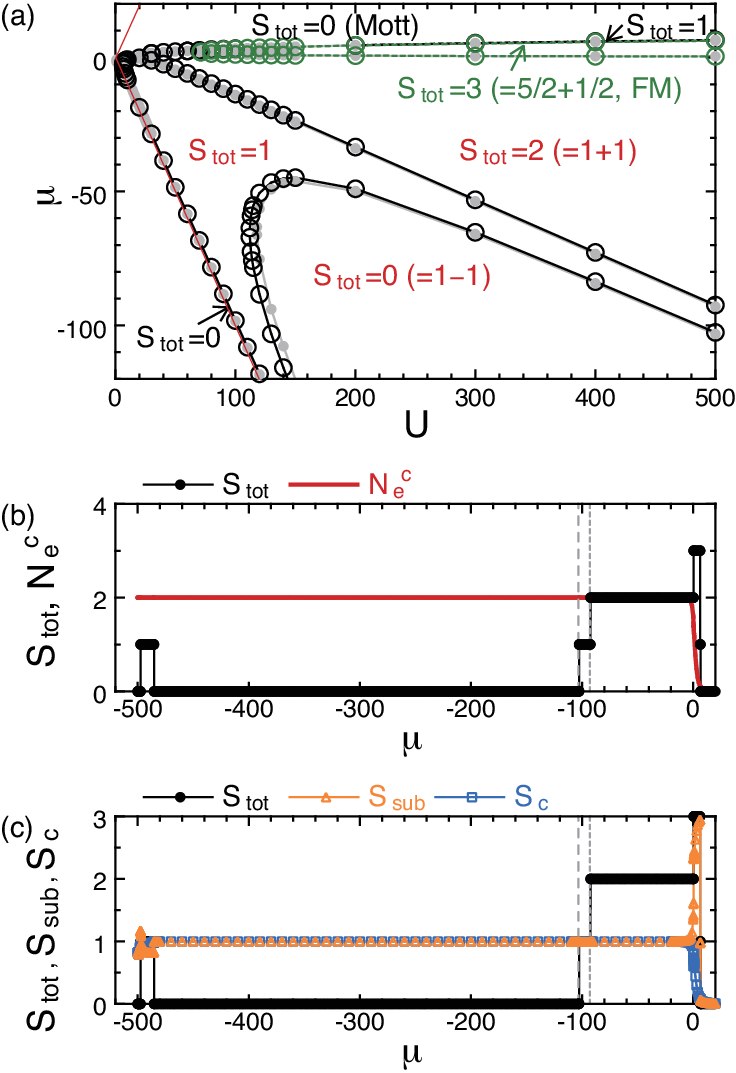}
\caption{
Properties of the effective $t$-$J$ model (\ref{Eq_Heff}) with $N=8$ and $N_{\mathrm{e}}=6$,
compared to the results of the Hubbard model
presented in Fig.~\ref{Fig_n8}.
(a) Ground-state phase diagram.
Open circles are results of the effective $t$-$J$ model,
and solid circles in gray are results of the Hubbard model,
given in Fig.~\ref{Fig_n8}(d).
Boundaries of the parameter space $\mu=\pm U$ are denoted by red lines.
(b) The total spin $S_{\mathrm{tot}}$
and the number of electrons in the center sites $N_{\mathrm{e}}^{\mathrm{c}}$
as a function of $\mu$ at $U=500$.
(c) The total spin $S_{\mathrm{tot}}$
together with those in the subsystem $S_{\mathrm{sub}}$ and in the center sites $S_{\mathrm{c}}$
as a function of $\mu$ at $U=500$.
In (b) and (c),
vertical dotted and dashed lines around $\mu \simeq -100$ denote
the transition point from $S_{\mathrm{tot}}=2$ to 1 and
that from $S_{\mathrm{tot}}=1$ to 0, respectively,
in the case of the Hubbard model.
}
\label{Fig_tJ}
\end{figure}

We note that we should consider carefully the effect of $\mu$
to have an appropriate effective $t$-$J$ model.
According to Eq.~(\ref{Eq_Jij}),
the exchange interaction is given by
$J_{\mathrm{ex}}^{\mathrm{sc}}=4t^2/U\{1-(\mu/U)^2\}$
for bonds connecting a site in the subsystem and a center site.
This quantity $J_{\mathrm{ex}}^{\mathrm{sc}}$ diverges at $U=\vert \mu \vert$,
which is the boundary of the parameter space of
$U>\vert \mu \vert$.
Considering these modifications,
we have
\begin{align}
  H_{\mathrm{eff}}^{\prime\prime}
  =&
  -t \sum_{\langle i,j \rangle,\sigma} (\tilde{c}_{i\sigma}^{\dag}\tilde{c}_{j\sigma} + \mathrm{H.c.})
  +\mu \sum_{i \in \mathrm{center}} \tilde{n}_{i}
\nonumber \\
  &
  +\sum_{\langle i,j \rangle}
  J_{ij} \left( \tilde{\bm{S}}_{i} \cdot \tilde{\bm{S}}_{j}-\frac{1}{4}\tilde{n}_{i}\tilde{n}_{j} \right),
\label{Eq_Heff_s2}
\end{align}
where $J_{ij}$ is given by Eq.~(\ref{Eq_Jij}).

In Fig.~\ref{Fig_tJ_simple2}(a),
we show the ground-state phase diagram.
We find again phases of the same set of values of $S_{\mathrm{tot}}$ as that of the Hubbard model (\ref{Eq_H}).
An improvement from
the case of the simplified model (\ref{Eq_Heff_s1}) is that
we find phase boundaries near $\mu=-U$
instead of vertical ones.
However, the difference of phase boundaries is still noticeable
when $\mu$ is large negative values.
In fact, for the states of $S_{\mathrm{tot}}=2$, 1, and 0
where $S_{\mathrm{sub}} \simeq 1$ and $S_{\mathrm{c}} \simeq 1$ in common,
the phase boundaries shift such that
the regions of the $S_{\mathrm{tot}}=2$ and $S_{\mathrm{tot}}=1$ states shrink
and the region of the $S_{\mathrm{tot}}=0$ state extends.

We find that
the dependences of
$N_{\mathrm{e}}^{\mathrm{c}}$, $S_{\mathrm{tot}}$, $S_{\mathrm{sub}}$, and $S_{\mathrm{c}}$
resemble those in the Hubbard model,
comparing Figs.~\ref{Fig_tJ_simple2}(b) and \ref{Fig_tJ_simple2}(c)
with Figs.~\ref{Fig_n8}(a) and \ref{Fig_n8}(b),
respectively.
However, the positions of transition points are located around $\mu \simeq -80$,
which are shifted from those in the Hubbard model around $\mu \simeq -100$ (vertical dotted and dashed lines).
This indicates a lack of a key ingredient for the quantitative agreements,
although properties of the magnetic states are described qualitatively.

We mention that
when $\mu$ gets close to $-U$,
the ground state becomes spin singlet,
which is similar to the case of the Hubbard model.
However,
$N_{\mathrm{e}}^{\mathrm{c}} \simeq 2$ keeps unchanged,
i.e., this transition is not caused by the double occupancy in the center sites,
contrary to the case of the Hubbard model where $N_{\mathrm{e}}^{\mathrm{c}} \simeq 4$.

\subsection{$t$-$J$ model with three-site pair-hopping}

Let us focus on the effective $t$-$J$ model,
fully described by Eq.~(\ref{Eq_Heff}) including the term of $J_{ijk}$.
In Fig.~\ref{Fig_tJ}(a),
we show the ground-state phase diagram.
Now the effective $t$-$J$ model reproduces the results of the Hubbard model fairly well.
Figures~\ref{Fig_tJ}(b) and \ref{Fig_tJ}(c) show the dependences of quantities,
which almost recover those of the Hubbard model
shown in Figs.~\ref{Fig_n8}(a) and \ref{Fig_n8}(b), respectively.
Indeed,
we cannot distinguish the results
from those of the Hubbard model
except that $N_{\mathrm{e}}^{\mathrm{c}} \simeq 2$ keeps unchanged near $\mu=-U$
due to the prohibition of the double occupancy
in the effective $t$-$J$ model.
These observations suggest that
we can discuss the property of the Hubbard model
making good use of the effective $t$-$J$ model.

\subsection{Classification of energies}

In the effective $t$-$J$ model,
we have the classification of different energy contributions,
such as potential, kinetic, and magnetic energies.
In particular,
we can classify the magnetic energies
which represent spin interactions
originally coming from the spin-independent Coulomb interaction in the Hubbard model.
To understand how various magnetic states appear
due to the competition among different energy contributions,
we classify the energies
within the effective $t$-$J$ model given by Eq.~(\ref{Eq_Heff}).
For this purpose,
we examine individual energy terms in the total energy $E$,
expressed as
\begin{equation}
  E = \langle H_{\mathrm{eff}} \rangle = E_{t}+E_{\mu}+E_{\mathrm{ex}}+E_{\mathrm{ph}}.
\end{equation}
Among these energy terms,
the potential energy,
\begin{equation}
  E_{\mu} =
  \mu \sum_{i \in \mathrm{center}} \langle \tilde{n}_{i} \rangle = \mu N_{\mathrm{e}}^{\mathrm{c}},
\label{Eq_Emu}
\end{equation}
gives the largest energy scale for negative $\mu$,
as we will see below.
The electron hopping term,
\begin{equation}
  E_{t} =
  -t \sum_{\langle i,j \rangle,\sigma} \langle (\tilde{c}_{i\sigma}^{\dag}\tilde{c}_{j\sigma} + \mathrm{H.c.}) \rangle,
\end{equation}
gives the kinetic energy.
The exchange interaction and the three-site pair-hopping terms,
\begin{equation}
  E_{\mathrm{ex}} =
  \sum_{\langle i,j \rangle}
  J_{ij} \left\langle \left( \tilde{\bm{S}}_{i} \cdot \tilde{\bm{S}}_{j}-\frac{1}{4}\tilde{n}_{i}\tilde{n}_{j} \right) \right\rangle,
\end{equation}
\begin{align}
  E_{\mathrm{ph}}
  =&
  -\sum_{\langle\langle i,j,k \rangle\rangle,\sigma}^{i \neq k}
  \frac{J_{ijk}}{4}
  \langle
  ( \tilde{c}_{i\sigma}^{\dag} \tilde{c}_{j,-\sigma}^{\dag} \tilde{c}_{j,-\sigma} \tilde{c}_{k\sigma}
\nonumber \\
  & \ \ \ \ \ \ \ \ \ \ \ \ \ \ \ \
    -\tilde{c}_{i,-\sigma}^{\dag} \tilde{c}_{j\sigma}^{\dag} \tilde{c}_{j,-\sigma} \tilde{c}_{k\sigma}
  + \mathrm{H.c.})
  \rangle,
\end{align}
respectively, give kinds of magnetic energies.

We further decompose each term to two parts:
One for the sum in the subsystem,
and the other for the sum over bonds involving the center sites.
Explicitly, for the electron hopping term, we have
\begin{equation}
  E_{t}^{\mathrm{sub}} =
  -t \sum_{\langle i,j \rangle\in\mathrm{subsystem},\sigma} \langle (\tilde{c}_{i\sigma}^{\dag}\tilde{c}_{j\sigma} + \mathrm{H.c.}) \rangle,
\end{equation}
\begin{equation}
  E_{t}^{\mathrm{c}} =
  E_{t}-E_{t}^{\mathrm{sub}}.
\end{equation}
For the exchange interaction term,
\begin{equation}
  E_{\mathrm{ex}}^{\mathrm{sub}} =
  \sum_{\langle i,j \rangle\in\mathrm{subsystem}}
  J_{ij} \left\langle \left( \tilde{\bm{S}}_{i} \cdot \tilde{\bm{S}}_{j}-\frac{1}{4}\tilde{n}_{i}\tilde{n}_{j} \right) \right\rangle,
\end{equation}
\begin{equation}
  E_{\mathrm{ex}}^{\mathrm{c}} =
  E_{\mathrm{ex}}-E_{\mathrm{ex}}^{\mathrm{sub}}.
\end{equation}
For the three-site pair-hopping term,
\begin{align}
  E_{\mathrm{ph}}^{\mathrm{sub}}
  =&
  -\sum_{\langle\langle i,j,k \rangle\rangle\in\mathrm{subsystem},\sigma}^{i \neq k}
  \frac{J_{ijk}}{4}
  \langle
  ( \tilde{c}_{i\sigma}^{\dag} \tilde{c}_{j,-\sigma}^{\dag} \tilde{c}_{j,-\sigma} \tilde{c}_{k\sigma}
\nonumber \\
  & \ \ \ \ \ \ \ \ \ \ \ \ \ \ \ \
   -\tilde{c}_{i,-\sigma}^{\dag} \tilde{c}_{j\sigma}^{\dag} \tilde{c}_{j,-\sigma} \tilde{c}_{k\sigma}
   + \mathrm{H.c.})
  \rangle,
\end{align}
\begin{equation}
  E_{\mathrm{ph}}^{\mathrm{c}} =
  E_{\mathrm{ph}}-E_{\mathrm{ph}}^{\mathrm{sub}}.
\end{equation}
Note that we do not decompose the potential energy $E_{\mu}$,
since it includes only the sum with respect to the center sites.

\begin{figure}[t] 
\centering
\includegraphics[clip,scale=0.65]{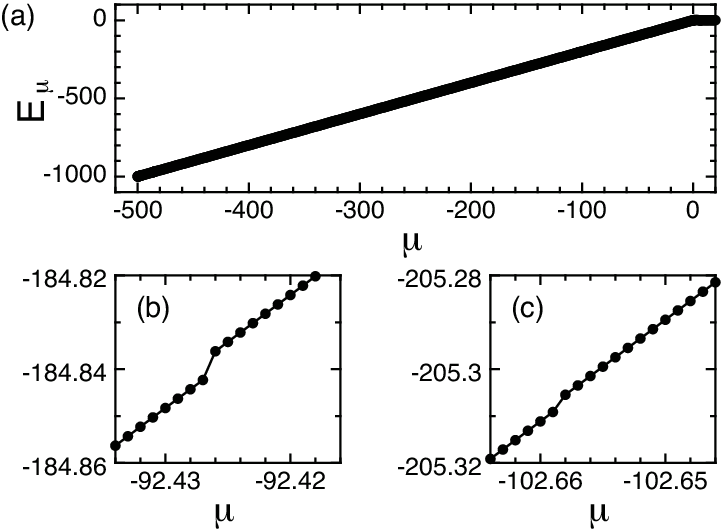}
\caption{
Part of ground-state energy of the effective $t$-$J$ model (\ref{Eq_Heff})
as a function of $\mu$ at $U=500$ with $N=8$ and $N_{\mathrm{e}}=6$.
(a) On-site potential energy term $E_{\mu}$.
In (b) and (c),
$E_{\mu}$ is plotted in magnified scales near the transition points around $\mu \simeq -100$.
}
\label{Fig_tJ_E_mu}
\end{figure}

In Fig.~\ref{Fig_tJ_E_mu},
we present the dependence of $E_{\mu}$ on $\mu$ at $U=500$.
We are particularly interested in the region of negative $\mu$,
where the center sites are half-filled
and $N_{\mathrm{e}}^{\mathrm{c}} \simeq 2$.
We anticipate that $E_{\mu} \simeq 2\mu$ from Eq.~(\ref{Eq_Emu}),
and it is actually observed in Fig.~\ref{Fig_tJ_E_mu}(a).
Assuming that $\mu$ takes large negative values,
the magnitude of $\mu$ is much larger than those of $t$, $J_{ij}$, and $J_{ijk}$,
so that $E_{\mu}$ gives the most contribution to the total energy.
We do not recognize any fine structures in the scale of Fig.~\ref{Fig_tJ_E_mu}(a).
However, if we take a closer look at the dependence near the transition points around $\mu \simeq -100$,
we find small jumps,
where $E_{\mu}$ suddenly drops with decreasing $\mu$,
as shown in Figs.~\ref{Fig_tJ_E_mu}(b) and \ref{Fig_tJ_E_mu}(c).
This suggests that these transitions around $\mu \simeq -100$ are accompanied by
a small sudden increase of $N_{\mathrm{e}}^{\mathrm{c}}$ toward $N^{\mathrm{c}}$ with decreasing $\mu$.
As we will see below,
the amplitudes of these jumps of $E_{\mu}$ are comparable with those of other energy terms,
i.e., $E_{t}$, $E_{\mathrm{ex}}$, and $E_{\mathrm{ph}}$.

\begin{figure}[t] 
\centering
\includegraphics[clip,scale=0.65]{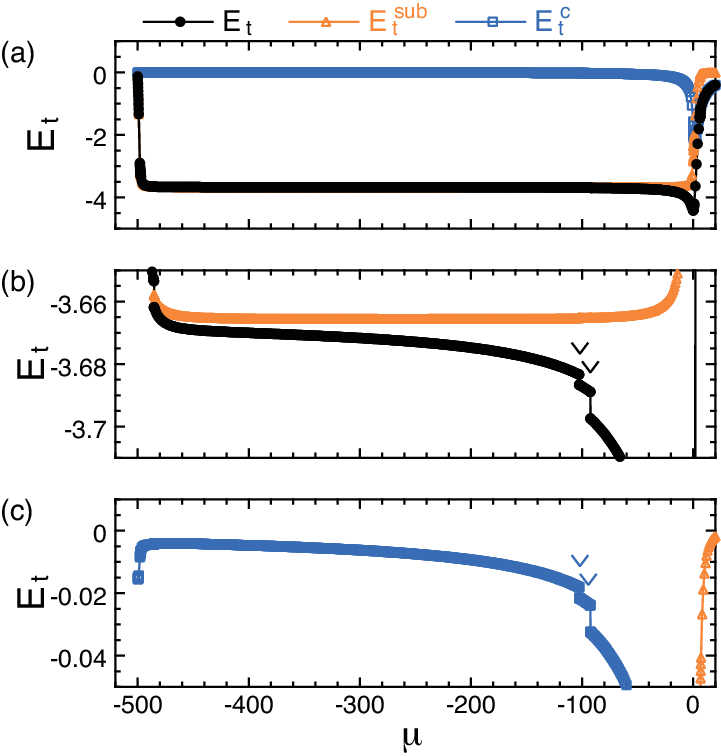}
\caption{
Part of ground-state energy of the effective $t$-$J$ model (\ref{Eq_Heff})
as a function of $\mu$ at $U=500$ with $N=8$ and $N_{\mathrm{e}}=6$.
(a) Electron hopping term $E_{t}$,
also plotted in magnified scales in (b) and (c).
Jumps at the transition points around $\mu \simeq -100$ are marked by arrowheads in (b) and (c).
}
\label{Fig_tJ_E_t}
\end{figure}

In Fig.~\ref{Fig_tJ_E_t},
we show the dependence of $E_{t}$.
It is natural that
$E_{t}^{\mathrm{sub}}$ dominates because electrons at the center sites hardly move
for negative $\mu$.
Indeed, we find that
$E_{t}^{\mathrm{sub}}$ has a large magnitude and gives the most contribution to $E_{t}$.
On the other hand,
$E_{t}^{\mathrm{c}}$ has a small magnitude
but gives the detailed dependence of $E_{t}$,
such as the structure of jumps at the transition points around $\mu \simeq -100$,
as marked in Figs.~\ref{Fig_tJ_E_t}(b) and \ref{Fig_tJ_E_t}(c).
As $\mu$ decreases to large negative values,
the magnitude of $E_{t}^{\mathrm{c}}$ becomes small little by little,
indicating that the electron motion through the center sites is suppressed.
This is because
the center sites approach half-filled asymptotically.

\begin{figure}[t] 
\centering
\includegraphics[clip,scale=0.65]{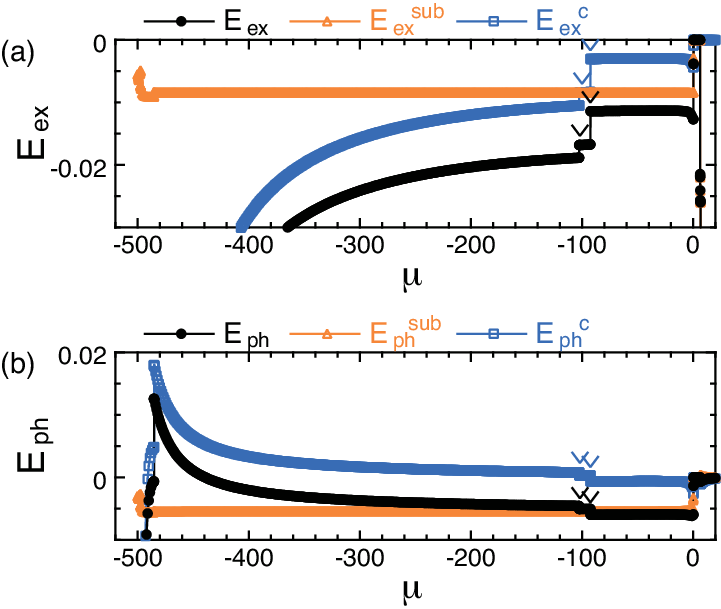}
\caption{
Part of ground-state energy of the effective $t$-$J$ model (\ref{Eq_Heff})
as a function of $\mu$ at $U=500$ with $N=8$ and $N_{\mathrm{e}}=6$.
(a) Exchange interaction term $E_{\mathrm{ex}}$
and
(b) three-site pair-hopping term $E_{\mathrm{ph}}$.
Jumps at the transition points around $\mu \simeq -100$ are marked by arrowheads.
}
\label{Fig_tJ_E_J}
\end{figure}

In Figs.~\ref{Fig_tJ_E_J}(a) and \ref{Fig_tJ_E_J}(b),
we show the dependences of $E_{\mathrm{ex}}$ and $E_{\mathrm{ph}}$, respectively.
Regarding $E_{\mathrm{ex}}$,
in the region of negative $\mu$,
$E_{\mathrm{ex}}^{\mathrm{sub}}$ is almost constant,
while $E_{\mathrm{ex}}^{\mathrm{c}}$ gives a main contribution to the $\mu$ dependence of $E_{\mathrm{ex}}$,
including the structure of jumps at the transition points around $\mu \simeq -100$,
as marked in Fig.~\ref{Fig_tJ_E_J}(a).
Such behaviors are also observed for $E_{\mathrm{ph}}$,
i.e.,
$E_{\mathrm{ph}}^{\mathrm{sub}}$ is almost constant,
while $E_{\mathrm{ph}}^{\mathrm{c}}$ gives a main contribution to the $\mu$ dependence of $E_{\mathrm{ph}}$.
We find that the magnitudes of $E_{\mathrm{ex}}$ and $E_{\mathrm{ph}}$ are rather small
in comparison with $E_{t}$,
indicating that the magnetic energy has a small energy scale
compared with the kinetic energy,
as is usually expected.

Here we note that
$E_{\mu}$, $E_{t}$, $E_{\mathrm{ex}}$, and $E_{\mathrm{ph}}$
show jumps with similar amplitudes
at the transition points around $\mu \simeq -100$,
even though they have different energy scales.
With decreasing $\mu$,
$E_{\mu}$ and $E_{\mathrm{ex}}$ decrease,
while $E_{t}$ and $E_{\mathrm{ph}}$ increase.
That is,
$E_{\mu}$ and $E_{\mathrm{ex}}$ give the energy gain,
while $E_{t}$ and $E_{\mathrm{ph}}$ cause the energy loss.
This indicates that
the transitions occur as a result of
the competition among different energy terms.

\begin{figure}[t] 
\centering
\includegraphics[clip,scale=0.65]{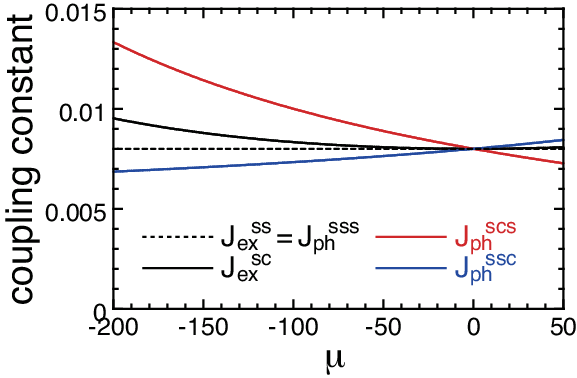}
\caption{
Coupling constants of magnetic interactions in the effective $t$-$J$ model
as a function of $\mu$ at $U=500$.
Note that all couplings are equivalent at $\mu=0$,
and $J_{\mathrm{ex}}^{\mathrm{ss}}$ and $J_{\mathrm{ph}}^{\mathrm{sss}}$ do not depend on $\mu$.
}
\label{Fig_Jexsc_Jphscs_Jphssc_U500}
\end{figure}

As mentioned above,
with decreasing $\mu$,
$E_{\mathrm{ex}}$ decreases, while $E_{\mathrm{ph}}$ increases,
i.e., different magnetic interactions compete with each other.
To clarify this behavior,
we discuss the property of the magnetic states
based on the analysis of the three-site pair-hopping
in a minimal three-site system (see Appendix B).
Since $E_{\mathrm{ex}}^{\mathrm{c}}$ and $E_{\mathrm{ph}}^{\mathrm{c}}$ cause
the structures of jumps of
$E_{\mathrm{ex}}$ [Fig.~\ref{Fig_tJ_E_J}(a)]
and
$E_{\mathrm{ph}}$ [Fig.~\ref{Fig_tJ_E_J}(b)],
respectively,
we take the three-site system including a center site.
In the negative $\mu$ region,
each center site catches almost one electron,
so that the three-site pair-hopping is relevant
for three sites with a center site in the middle [Fig.~\ref{Fig_3-site-config}(b)].
We focus on this three-site configuration in the following discussion.
In the subspace of two electrons with antiparallel spins,
we have four basis states,
and the eigenenergies are given by diagonalizing the $4 \times 4$ Hamiltonian matrix.
The lowest eigenenergy is $\lambda_{1} = -J_{\mathrm{ph}}^{\mathrm{scs}}/2$
and the eigenstate is
\begin{align}
  \vert \phi_{1} \rangle
  =&
  \frac{1}{2}
  (\vert 0\rangle_{i} \vert \! \uparrow\rangle_{j} \vert \! \downarrow\rangle_{k}
  -\vert \! \downarrow\rangle_{i} \vert \! \uparrow\rangle_{j} \vert 0\rangle_{k}
\nonumber \\
  &
  -\vert 0\rangle_{i} \vert \! \downarrow\rangle_{j} \vert \! \uparrow\rangle_{k}
  +\vert \! \uparrow\rangle_{i} \vert \! \downarrow\rangle_{j} \vert 0\rangle_{k})
\nonumber \\
  =&
  \frac{1}{\sqrt{2}}( \vert 0\rangle_{i} \vert s \rangle_{jk} + \vert s \rangle_{ij} \vert 0\rangle_{k}),
\label{Eq_phi1}
\end{align}
where
$\vert s \rangle_{ij} =
\frac{1}{\sqrt{2}}
( \vert \! \uparrow\rangle_{i} \vert \! \downarrow\rangle_{j}
- \vert \! \downarrow\rangle_{i} \vert \! \uparrow\rangle_{j} )$
is the spin singlet state on the bond $ij$.
Thus, the three-site pair-hopping stabilizes the spin singlet states at a local level.
We note that
$\vert \phi_{1} \rangle$ is also the eigenstate of the exchange interaction
with the eigenenergy $-J_{\mathrm{ex}}^{\mathrm{sc}}$,
and the exchange interaction stabilizes the spin singlet states locally.
We may expect that
$\vert \phi_{1} \rangle$ would contribute to the magnetic states in the negative $\mu$ region.
However,
in the state $\vert \phi_{1} \rangle$,
eigenenergies $-J_{\mathrm{ex}}^{\mathrm{sc}}$ and $-J_{\mathrm{ph}}^{\mathrm{scs}}/2$
decrease with decreasing $\mu$,
since $J_{\mathrm{ex}}^{\mathrm{sc}}$ and $J_{\mathrm{ph}}^{\mathrm{scs}}$ are positive
and increase with decreasing $\mu$,
as seen in Fig.~\ref{Fig_Jexsc_Jphscs_Jphssc_U500}.
The dependence of $-J_{\mathrm{ex}}^{\mathrm{sc}}$ agrees with that of $E_{\mathrm{ex}}$,
whereas the dependence of $-J_{\mathrm{ph}}^{\mathrm{scs}}/2$ contradicts that of $E_{\mathrm{ph}}$.

On the other hand,
the highest eigenenergy is $\lambda_{4} = J_{\mathrm{ph}}^{\mathrm{scs}}/2$
and the eigenstate is
\begin{align}
  \vert \phi_{4} \rangle
  =&
  \frac{1}{2}
  (\vert 0\rangle_{i} \vert \! \uparrow\rangle_{j} \vert \! \downarrow\rangle_{k}
  +\vert \! \downarrow\rangle_{i} \vert \! \uparrow\rangle_{j} \vert 0\rangle_{k}
\nonumber \\
  &
  -\vert 0\rangle_{i} \vert \! \downarrow\rangle_{j} \vert \! \uparrow\rangle_{k}
  -\vert \! \uparrow\rangle_{i} \vert \! \downarrow\rangle_{j} \vert 0\rangle_{k})
\nonumber \\
  =&
  \frac{1}{\sqrt{2}}( \vert 0\rangle_{i} \vert s \rangle_{jk} - \vert s \rangle_{ij} \vert 0\rangle_{k}),
\label{Eq_phi4}
\end{align}
which is also the superposition of the local spin singlet states,
while signs of coefficients are different from the case of $\vert \phi_{1} \rangle$.
$\vert \phi_{4} \rangle$ is also the eigenstate of the exchange interaction
with the eigenenergy $-J_{\mathrm{ex}}^{\mathrm{sc}}$.
In the state $\vert \phi_{4} \rangle$,
$-J_{\mathrm{ex}}^{\mathrm{sc}}$ decreases and $J_{\mathrm{ph}}^{\mathrm{scs}}/2$ increases
with decreasing $\mu$,
which coincides with the dependences of $E_{\mathrm{ex}}$ and $E_{\mathrm{ph}}$,
respectively.
In the light of this coincidence,
we would attribute the contribution of $\vert \phi_{4} \rangle$ as follows.
Considering that
$S_{\mathrm{sub}} \simeq 1$ is stable for negative $\mu$,
we suppose that the wavefunction is influenced by the occurrence of the spin polarization in the subsystem.
To realize the spin polarization in the subsystem,
a bonding state would be formed in the subsystem,
where basis states caused by the hole motion in the subsystem are summed with the same sign
to construct the wavefunction.
In
$\vert \phi_{4} \rangle$
given by Eq.~(\ref{Eq_phi4}),
such symmetric summations are taken over the basis states of the subsystem.
That is,
$\vert 0\rangle_{i} \vert \! \uparrow\rangle_{j} \vert \! \downarrow\rangle_{k}$
and
$\vert \! \downarrow\rangle_{i} \vert \! \uparrow\rangle_{j} \vert 0\rangle_{k}$
are summed with the same positive sign, and
$\vert 0\rangle_{i} \vert \! \downarrow\rangle_{j} \vert \! \uparrow\rangle_{k}$
and
$\vert \! \uparrow\rangle_{i} \vert \! \downarrow\rangle_{j} \vert 0\rangle_{k}$
are summed with the same negative sign.
In contrast,
in
$\vert \phi_{1} \rangle$
given by Eq.~(\ref{Eq_phi1}),
antisymmetric summations are taken.
Thus, we conclude that
$\vert \phi_{4} \rangle$
would contribute for the spin polarization in the subsystem.

\section{EXTENDED LATTICES WITH MORE THAN TWO UNITS}

\begin{figure}[t] 
\centering
\includegraphics[clip,scale=0.65]{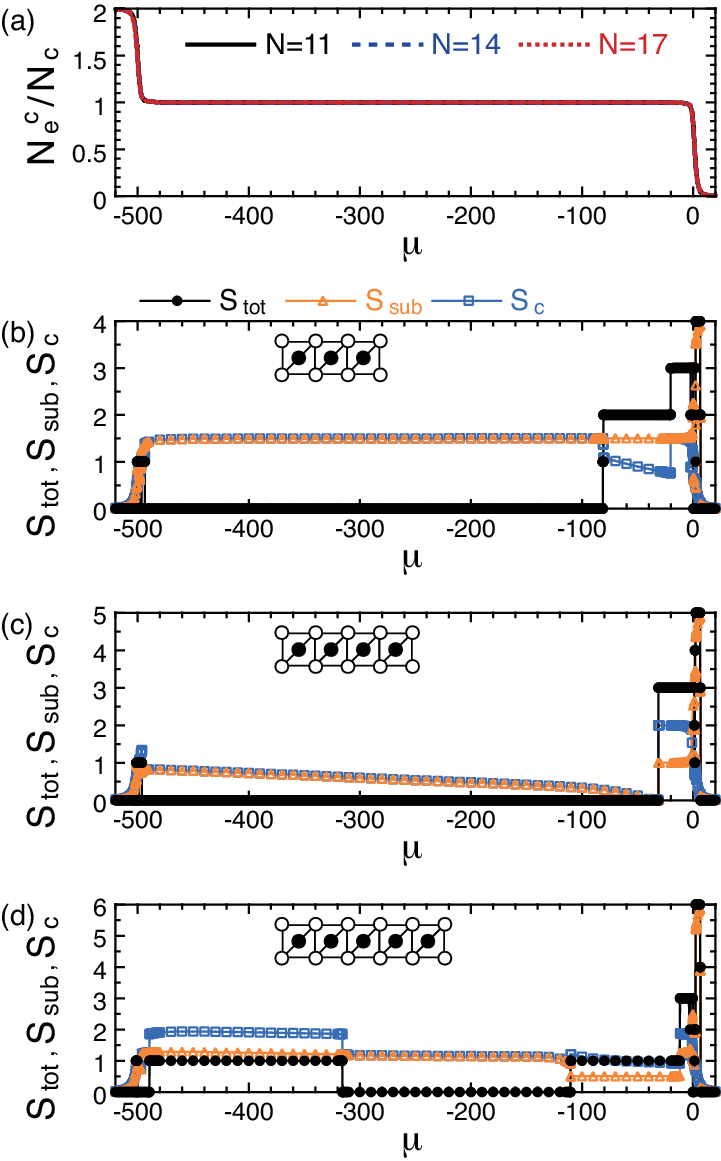}
\caption{
Properties of the Hubbard model (\ref{Eq_H}) with $N=11$, 14, and 17.
(a) The electron density in the center sites $N_{\mathrm{e}}^{\mathrm{c}}/N_{\mathrm{c}}$
as a function of $\mu$ at $U=500$.
Note that data of $N_{\mathrm{e}}^{\mathrm{c}}/N_{\mathrm{c}}$ for several system sizes collapse.
The total spin $S_{\mathrm{tot}}$
and those in the subsystem $S_{\mathrm{sub}}$ and in the center sites $S_{\mathrm{c}}$
as a function of $\mu$ at $U=500$
for the systems of
(b) $N=11$,
(c) $N=14$,
and
(d) $N=17$.
}
\label{Fig_n11-14-17}
\end{figure}

Thus far, we have studied the system with $N=8$ sites, composed of two units.
In this section,
we move to the extended lattices with more than two units.
In Fig.~\ref{Fig_n11-14-17},
we present numerical results of quantities as a function of $\mu$ at $U=500$
for $N=11$ (three units), $N=14$ (four units), and $N=17$ (five units).
In Figs.~\ref{Fig_n11-14-17}(b)-(d),
the saturated FM state is found in
$2 \lesssim \mu \lesssim 8$
(extended Nagaoka ferromagnetism).
Below this region,
electrons move around the whole system
because of the small $\mu$,
and the electron distribution sharply changes with $\mu$,
as shown in Fig.~\ref{Fig_n11-14-17}(a).
The total spin changes according to the sharp change of the electron distribution,
which causes a complicated dependence near $\mu=0$,
as shown in Figs.~\ref{Fig_n11-14-17}(b)-(d).

\begin{figure}[t] 
\centering
\includegraphics[clip,scale=0.65]{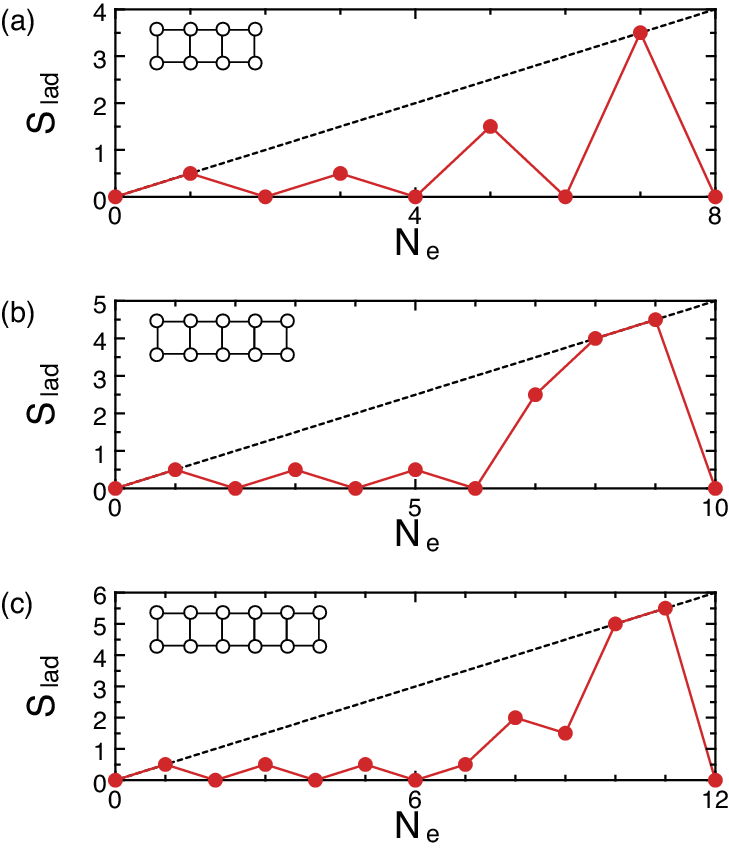}
\caption{
The total spin of the two-leg ladder system $S_{\mathrm{lad}}$
in the limit of infinite $U$ as a function of the number of electrons $N_{\mathrm{e}}$ for
(a) $N_{\mathrm{lad}}=8$,
(b) $N_{\mathrm{lad}}=10$,
and
(c) $N_{\mathrm{lad}}=12$.
}
\label{Fig_two-leg-ladder_n8-10-12}
\end{figure}

In the negative $\mu$ region,
the center sites are half-filled until $\mu$ goes down to near $-U$
regardless of the system size,
as shown in Fig.~\ref{Fig_n11-14-17}(a).
However,
the dependences of the spin states look rather different for different system sizes as follows.

For $N=11$,
in the negative $\mu$ region,
three holes are doped into the subsystem of $N_{\mathrm{sub}}=8$.
In Fig.~\ref{Fig_n11-14-17}(b),
we see that $S_{\mathrm{tot}}$ decreases sequentially one by one from three to zero with decreasing $\mu$.
In these states of $S_{\mathrm{tot}}=3$, 2, 1, and 0,
$S_{\mathrm{sub}} \simeq 3/2$ is stable in common.
On the other hand,
$S_{\mathrm{c}}$ changes with decreasing $\mu$.
That is,
$S_{\mathrm{c}} \simeq 3/2$ in the $S_{\mathrm{tot}}=3$ state,
it turns to take non-half-integer real values in the $S_{\mathrm{tot}}=2$ state,
it shows a jump and takes non-half-integer real values in the $S_{\mathrm{tot}}=1$ state,
and $S_{\mathrm{c}} \simeq 3/2$ again in the $S_{\mathrm{tot}}=0$ state.
Here we consider the state of the subsystem
based on the two-leg ladder system of $N_{\mathrm{lad}}=8$,
in the same manner as the case of $N=8$ and $N_{\mathrm{lad}}=6$ discussed in Sec.~III~B.
As shown in Fig.~\ref{Fig_two-leg-ladder_n8-10-12}(a),
a partial FM state of $S_{\mathrm{lad}}=3/2$ appears for $N_{\mathrm{e}}=5$, i.e., three holes.
Thus,
the occurrence of stable $S_{\mathrm{sub}} \simeq 3/2$
would be attributed to the fact that
the two-leg ladder system with three holes has $S_{\mathrm{lad}}=3/2$.

For $N=14$,
in the negative $\mu$ region,
four holes are doped into the subsystem of $N_{\mathrm{sub}}=10$.
In Fig.~\ref{Fig_n11-14-17}(c),
we find that $S_{\mathrm{tot}}$ jumps from three to zero without taking intermediate values
with decreasing $\mu$.
There,
$S_{\mathrm{sub}} \simeq 1$ and $S_{\mathrm{c}} \simeq 2$ in the $S_{\mathrm{tot}}=3$ state,
while $S_{\mathrm{sub}}$ and $S_{\mathrm{c}}$ becomes almost zero simultaneously
at the transition to the $S_{\mathrm{tot}}=0$ state.
Within the $S_{\mathrm{tot}}=0$ state,
$S_{\mathrm{sub}}$ and $S_{\mathrm{c}}$ coincide with each other.
They gradually increase taking non-integer real values.
On the other hand,
the ground state of the two-leg ladder system with $N_{\mathrm{lad}}=10$ and $N_{\mathrm{e}}=6$ (four holes)
has $S_{\mathrm{lad}}=0$,
as shown in Fig.~\ref{Fig_two-leg-ladder_n8-10-12}(b).
Comparing $S_{\mathrm{sub}}$ with $S_{\mathrm{lad}}$,
$S_{\mathrm{sub}} \simeq 0$,
which occurs just below the phase boundary between $S_{\mathrm{tot}}=3$ and $S_{\mathrm{tot}}=0$,
agrees with $S_{\mathrm{lad}}=0$.
However,
except for this agreement,
$S_{\mathrm{sub}}$ deviates from zero,
which disagrees with $S_{\mathrm{lad}}=0$.
From these observations,
we consider that
the transition from $S_{\mathrm{tot}}=3$ to $S_{\mathrm{tot}}=0$ emerges
to realize $S_{\mathrm{sub}} \simeq 0$,
since it corresponds to $S_{\mathrm{lad}}=0$.
In contrast,
$S_{\mathrm{sub}} \simeq 1$ in the $S_{\mathrm{tot}}=3$ state is not realized as the lowest-energy state of the subsystem,
but as an excited state
induced by the correlation between the subsystem and the center sites.
The same scenario also applies to the deviation of $S_{\mathrm{sub}}$ in the $S_{\mathrm{tot}}=0$ state,
i.e.,
finite $S_{\mathrm{sub}}$ is induced by the correlation between the subsystem and the center sites.

For $N=17$,
in the negative $\mu$ region,
five holes are doped into the subsystem of $N_{\mathrm{sub}}=12$.
In Fig.~\ref{Fig_n11-14-17}(d),
we find that with decreasing $\mu$,
$S_{\mathrm{tot}}$ decreases as three, one, zero,
and, after some interval of $\mu$,
increases to one again.
Among these states,
in the $S_{\mathrm{tot}}=1$ state with larger $\mu$
($-110.3 \lesssim \mu \lesssim -12.0$),
$S_{\mathrm{sub}} \simeq 1/2$ and $S_{\mathrm{c}} \simeq 1$.
The stable $S_{\mathrm{sub}} \simeq 1/2$ is reasonably understood
as the formation of the lowest-energy state in the subsystem,
since the two-leg ladder system of $N_{\mathrm{lad}}=12$ and $N_{\mathrm{e}}=7$ (five holes)
gives $S_{\mathrm{lad}}=1/2$,
as shown in Fig.~\ref{Fig_two-leg-ladder_n8-10-12}(c).
We note that there are five electrons in the center sites
and $S_{\mathrm{c}}$ is supposed to be half-integer
if the center sites are independent from the subsystem,
but actually $S_{\mathrm{c}} \simeq 1$,
which is caused by the correlation between the subsystem and the center sites.
On the other hand,
in the $S_{\mathrm{tot}}=0$ state,
$S_{\mathrm{sub}}$ and $S_{c}$ coincide with each other and they are slightly larger than one,
while
in the $S_{\mathrm{tot}}=1$ state with smaller $\mu$ ($-489.4 \lesssim \mu \lesssim -315.7$),
$S_{\mathrm{sub}}$ is slightly larger than one,
and $S_{c}$ is slightly smaller than two.
Thus,
$S_{\mathrm{sub}}$ and $S_{c}$ do not take half-integer values,
although there are odd numbers of electrons in both the subsystem and the center sites,
indicating that the subsystem and the center sites are highly correlated.

These results indicate that partial FM states occur in the negative $\mu$ region.
However, it is difficult to extrapolate the present results to the large system systematically,
and to conclude what the state is in the thermodynamic limit.

\begin{figure}[t] 
\centering
\includegraphics[clip,scale=0.65]{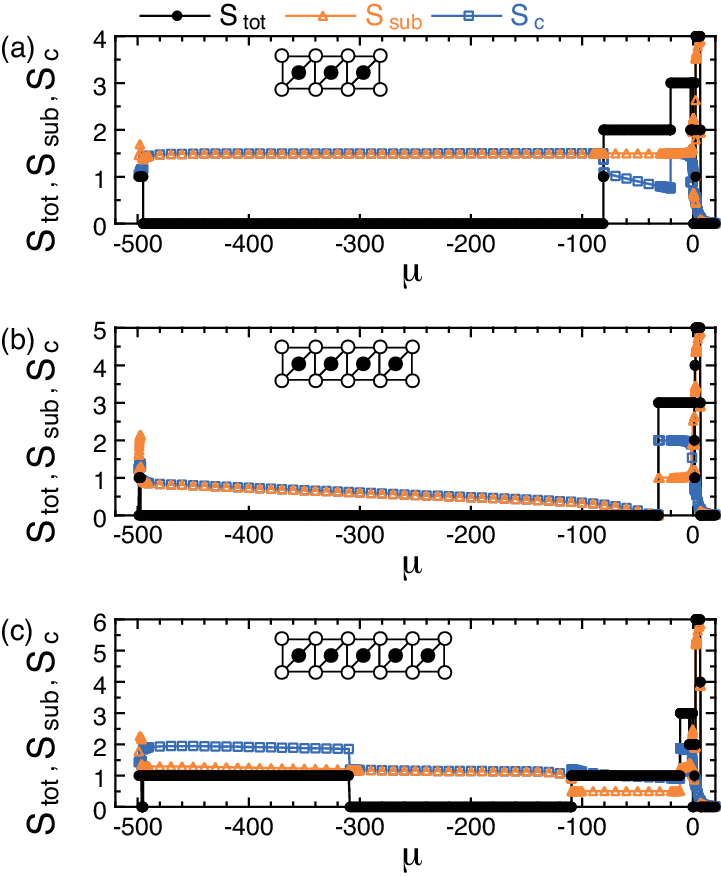}
\caption{
Properties of the effective $t$-$J$ model (\ref{Eq_Heff}) for several system sizes.
The total spin $S_{\mathrm{tot}}$
and those in the subsystem $S_{\mathrm{sub}}$ and in the center sites $S_{\mathrm{c}}$
as a function of $\mu$ at $U=500$
for the systems of
(a) $N=11$,
(b) $N=14$,
and
(c) $N=17$.
}
\label{Fig_tJ_n11-14-17}
\end{figure}

Finally,
we investigate properties of the effective $t$-$J$ model (\ref{Eq_Heff})
for several system sizes,
as shown in Fig.~\ref{Fig_tJ_n11-14-17}.
Compared to the results of the Hubbard model (\ref{Eq_H}) given in Fig.~\ref{Fig_n11-14-17},
we see that the dependences obtained by the effective $t$-$J$ model well reproduce
those of the Hubbard model
including the positions of the transition points,
indicating the validity of the effective $t$-$J$ model,
except for the region close to $\mu=-U$.
It is useful to use the effective $t$-$J$ model
to understand microscopic properties of various magnetic states.
However,
a detailed in-depth discussion for large system sizes is beyond the scope of the present paper.

\section{SUMMARY AND DISCUSSION}

In the present paper
we investigated magnetic ground states of the Hubbard model
including sites that work as the particle bath,
which we have introduced for the extended Nagaoka ferromagnetism
\cite{Miyashita2008,Onishi2014,Onishi2022,Onishi2023}.
Previously we have studied the extended Nagaoka ferromagnetism
in the weakly doped region.
In the present work,
we focused on the highly doped region
where the extended Nagaoka ferromagnetism disappears
due to the excess hole doping.

We performed detailed analyses for the system with $N=8$ sites
as a minimal case of the extended lattices.
In the highly doped region,
each center site catches one electron,
and the number of electrons in the center sites $N_{\mathrm{e}}^{\mathrm{c}} \simeq 2$ is almost fixed.
The total spin $S_{\mathrm{tot}}$ changes as 2, 1, and 0
with decreasing $\mu$,
while the total spin in the subsystem $S_{\mathrm{sub}} \simeq 1$
and that in the center sites $S_{\mathrm{c}} \simeq 1$
are stable in common.
That is, four electrons in the subsystem partially polarize.
To understand the mechanism,
we discussed the state of the subsystem
by referring to the ground state of the two-leg ladder system
that corresponds to the part of the subsystem excluding the center sites.
The total spin of the two-leg ladder system with four electrons is
$S_{\mathrm{lad}}=1$,
which coincides with $S_{\mathrm{sub}} \simeq 1$.
Thus,
we attribute the properties of the state of the subsystem
to those of the ground state in the two-leg ladder system.
We note that the hole doping in the subsystem is controlled continuously
by the chemical potential of the center sites.
The present scheme provides a unique approach for studying the doping effect
from the viewpoint of gradual hole doping.

However,
we found that for larger system sizes,
other kinds of magnetic states which are not the ground state but excited states of the two-leg ladder system
are also induced
by the correlation between the subsystem and the center sites.
This causes the complexity of the doping effect,
which does not arise in merely the subsystem.
In other words,
by preparing a well-designed system of the subsystem and the center sites,
we can realize not only the ground state
but also a variety of excited states
that are not accessible if we consider only the subsystem.
We expect that
this prospect would give a new pathway
to explore novel quantum states and functionalities of strongly correlated electron systems.

Moreover, to gain insight into the microscopic origin of the magnetic phase diagram,
we studied the ground state of the effective $t$-$J$ model,
derived from the Hubbard model
by considering the second-order processes of the electron hopping.
The ground-state phase diagram of the Hubbard model is well reproduced by the effective $t$-$J$ model (\ref{Eq_Heff}),
whereas the simplified $t$-$J$ models (\ref{Eq_Heff_s1}) and (\ref{Eq_Heff_s2}) are insufficient.
The three-site pair-hopping is required for the quantitative agreements,
although this term is often discarded in usual treatments of the $t$-$J$ model.

Based on the effective $t$-$J$ model,
we investigated the classification of different energy terms,
such as potential, kinetic, and magnetic energies.
These energy terms have different energy scales,
but they show jumps of similar amplitudes at the transition points,
indicating that the competition among different energy terms plays a role in the transitions.
In particular,
for the magnetic energies,
the exchange interaction term decreases
with decreasing $\mu$,
while the three-site pair-hopping term increases,
i.e.,
they change with $\mu$ in the opposite signs.
We argued that
this behavior would be attributed to the occurrence of the spin polarization in the subsystem.
To realize the spin polarization in the subsystem,
a bonding state would be formed in the subsystem.
This causes a high-energy state for the three-site pair-hopping
and the $\mu$ dependence of the three-site pair-hopping term opposite to the exchange interaction term.

We envisage that the present mechanism for doping control can be realized
by using advanced techniques to manipulate functional structures and generate quantum simulators,
such as cold atoms in optical lattices~\cite{Bloch2005,Bloch2012,Gross2017},
molecular magnets~\cite{Verdaguer1999,Coronado2013,Coronado2019},
quantum dots~\cite{Hensgens2017,Dehollain2020,Thakur2023},
and superconducting circuits~\cite{Houck2012}.
By preparing the present setup that includes the sites of the particle bath with a controllable way,
we can manipulate the total spin.
The dynamical aspect of this type of spin state manipulation is also accessible,
which would be relevant for developing spintronic devices or spin qubits.
This perspective is an interesting direction of future research.


\begin{acknowledgments}
Computations were performed on supercomputers
at the Japan Atomic Energy Agency
and at the Institute for Solid State Physics, the University of Tokyo.
This work was in part supported by JSPS KAKENHI Grant Numbers JP23K03331 and JP24K01332.
\end{acknowledgments}

\appendix
\section{Technical remarks on Lanczos calculations}

The Lanczos method is an iterative power method,
in which we operate the Hamiltonian matrix to a Lanczos vector
to generate an orthogonal basis set in the Krylov subspace
and tridiagonalize the Hamiltonian matrix.
To check the convergence of the resultant lowest eigenenergy,
we measure the difference of the lowest eigenenergy from the previous Lanczos step,
\begin{equation}
  \Delta E = |(E_{i}-E_{i-1})/E_{i-1}|,
\end{equation}
where $E_{i}$ is the lowest eigenenergy obtained in the $i$-th Lanczos step.
As an indicator for the convergence of the eigenvector,
we also measure the residual,
\begin{equation}
  r = || H |\psi\rangle_{i} - E_{i} |\psi\rangle_{i} ||,
\end{equation}
where $|\psi\rangle_{i}$ is the lowest eigenvector in the $i$-th Lanczos step.
After an appropriate $M$ Lanczos steps,
we compute the lowest eigenenergy at every Lanczos step
and check if $\Delta E$ becomes below a specified small value,
$10^{-10}$ in the present calculations.
We can obtain a good convergence
with a much smaller number of Lanczos steps than the dimension of the Hamiltonian matrix.

However,
we often encounter slow convergence
when there are nearly degenerate low-energy states.
This type of difficulty occurs
near transition points and also in the negative $\mu$ region in the present model.
For negative $\mu$,
electrons come to the center sites and holes are doped into the subsystem.
Thus, both charge and spin degrees of freedom are active,
which causes coexisting different energy scales relevant to the low-energy states.
In such a multiscale situation,
nearly degenerate low-energy states appear
due to a subtle balance among different energy terms.

\begin{figure}[t] 
\centering
\includegraphics[clip,scale=0.65]{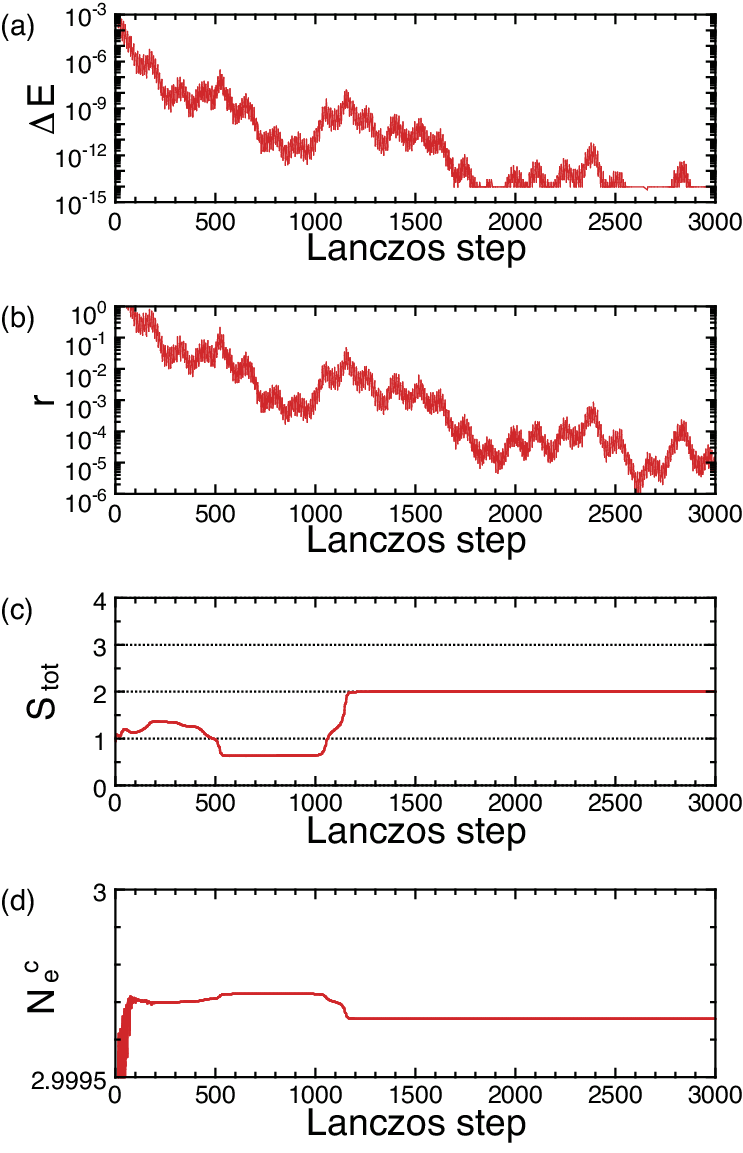}
\caption{
Lanczos results for $N=11$, $U=500$, and $\mu=-80$.
(a) The difference of the lowest eigenenergy from the previous Lanczos step $\Delta E$,
(b) the residual $r$,
(c) the total spin $S_{\mathrm{tot}}$, and
(d) the number of electrons in the center sites $N_{\mathrm{e}}^{\mathrm{c}}$
as a function of the Lanczos step.
}
\label{Fig_lanczos_n11}
\end{figure}

\begin{figure}[t] 
\centering
\includegraphics[clip,scale=0.65]{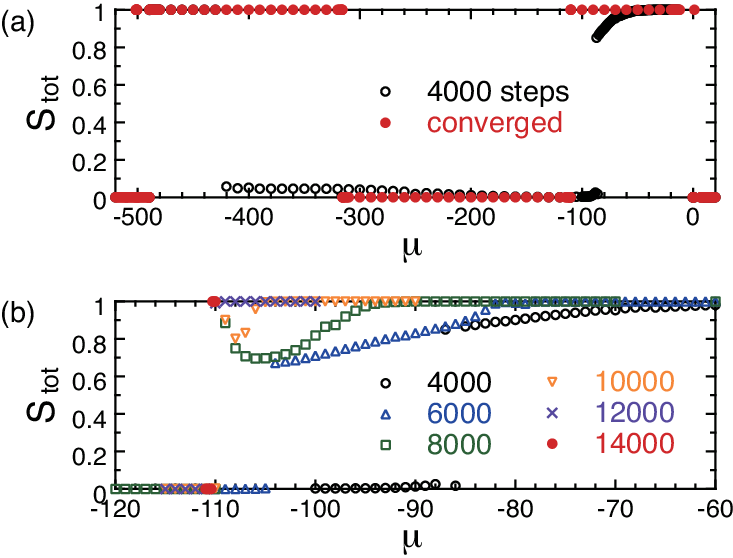}
\caption{
Lanczos results for $N=17$ and $U=500$.
(a) The total spin $S_{\mathrm{tot}}$ for $M=4000$ Lanczos steps and the converged value.
(b) $S_{\mathrm{tot}}$ for various $M$,
demonstrating how we attain the convergence with increasing $M$.
}
\label{Fig_lanczos_n17}
\end{figure}

As a typical example,
Figure~\ref{Fig_lanczos_n11} shows the dependences of quantities on the Lanczos step
in the case of $N=11$, $U=500$, and $\mu=-80$.
In Fig.~\ref{Fig_lanczos_n11}(a),
we find that $\Delta E$ becomes as small as $10^{-10}$ at around 600 Lanczos steps,
which indicates that the ground-state energy would be obtained accurately.
Moreover, as shown in Fig.~\ref{Fig_lanczos_n11}(b),
$r$ is evaluated to be $3.6\times 10^{-3}$ there,
which suggests that the eigenvector would also be obtained with a certain accuracy.
However,
as shown in Fig.~\ref{Fig_lanczos_n11}(c),
the total spin takes a non-integer value at around 600 Lanczos steps,
and it remains stable until around 1000 Lanczos steps.
When we continue Lanczos iterations further,
the total spin eventually converges to 2.
As shown in Fig.~\ref{Fig_lanczos_n11}(d),
the number of electrons in the center sites also shows deviation
and convergence in conjunction with the total spin.
These findings demonstrate that
the accuracy of the eigenvector is insufficient to give physical quantities correctly,
even when we obtain the ground-state energy within a required accuracy.
Here we note that,
because of nearly degenerate low-energy states,
it is difficult to apply the inverse iteration method effectively
to improve the accuracy of the eigenvector.
In the present case,
the total spin should be integer,
and this fact can be used for the check of the convergency.
Practically,
we perform Lanczos iterations
until the total spin converges to an integer value within five digits.

With increasing the system size,
the quasi-degeneracy of the low-energy states becomes severe,
and the number of Lanczos steps required for the convergence increases.
In Fig.~\ref{Fig_lanczos_n17},
we demonstrate how the total spin converges with the Lanczos iterations
in the case of $N=17$.
In Fig.~\ref{Fig_lanczos_n17}(a),
we see that for $M=4000$,
with decreasing $\mu$,
the total spin starts to decrease from 1 around $\mu=-60$,
and jumps to nearly zero at around $\mu=-85$.
Below this jump, the total spin seems to take zero in some interval of $\mu$,
but it is still within two digits,
indicating that the accuracy is insufficient.
Moreover, it gradually increases with decreasing $\mu$ down to around $\mu=-420$.
As shown in Fig.~\ref{Fig_lanczos_n17}(b),
as $M$ increases,
the total spin converges to integer values
and the transition point
between the states of $S_{\mathrm{tot}}=1$ and $S_{\mathrm{tot}}=0$
is precisely determined.

We emphasize that we are lucky to have the total spin which must be multiples of $1/2$
due to the spin SU(2) symmetry.
If the evaluated value does not satisfy such a strict condition,
we know that the Lanczos iteration is insufficient.
Without such a good guideline,
it is difficult to judge whether the Lanczos iteration is enough or not.
Thus, it is recommended to observe some quantities to check the convergence.

\section{Three-site pair-hopping in $t$-$J$ model}

In this Appendix,
we give a brief explanation about the property of the three-site pair-hopping.
Here we consider a minimal three-site system
of neighboring three sites $\langle\langle i,j,k \rangle\rangle$,
described by
\begin{align}
  H_{\mathrm{ph}}
  =&
  -\sum_{\sigma}
  \frac{J_{ijk}}{4}
  ( \tilde{c}_{i\sigma}^{\dag} \tilde{c}_{j,-\sigma}^{\dag} \tilde{c}_{j,-\sigma} \tilde{c}_{k\sigma}
\nonumber \\
  & \ \ \ \ \ \ \ \ \ \ \ \ \ \ \ \
    -\tilde{c}_{i,-\sigma}^{\dag} \tilde{c}_{j\sigma}^{\dag} \tilde{c}_{j,-\sigma} \tilde{c}_{k\sigma}
  + \mathrm{H.c.}),
\end{align}
where $J_{ijk}$ is given by Eq.~(\ref{Eq_Jijk}).
Note that the three-site pair-hopping is effective
when a hole is present adjacent to a pair of electrons with antiparallel spins,
as depicted in Fig.~\ref{Fig_tJ_process}(b).
Thus,
in the subspace of two electrons with opposite spins,
we have four basis states,
\begin{equation}
\left\{
  \begin{aligned}
  & \vert 0\rangle_{i} \, \vert \! \uparrow\rangle_{j} \, \vert \! \downarrow\rangle_{k}, \\
  & \vert \! \downarrow\rangle_{i} \, \vert \! \uparrow\rangle_{j} \, \vert 0\rangle_{k}, \\
  & \vert 0\rangle_{i} \, \vert \! \downarrow\rangle_{j} \, \vert \! \uparrow\rangle_{k}, \\
  & \vert \! \uparrow\rangle_{i} \, \vert \! \downarrow\rangle_{j} \, \vert 0\rangle_{k},
  \end{aligned}
\right.
\end{equation}
and $H_{\mathrm{ph}}$ is represented by the $4 \times 4$ matrix,
\begin{equation}
  H_{\mathrm{ph}}=
  -\frac{J_{ijk}}{4}
  \left(
  \begin{array}{cccc}
  0 & -1 & 0 & 1 \\
  -1 & 0 & 1 & 0 \\
  0 & 1 & 0 & -1 \\
  1 & 0 & -1 & 1
  \end{array}
  \right).
\end{equation}
By diagonalizing the $4 \times 4$ matrix,
eigenenergies
$\{ \lambda_{n} \}$
and
eigenstates
$\{ \vert \phi_{n} \rangle \}$
are obtained as
\begin{align}
  \lambda_{1} = -\frac{J_{ijk}}{2},
  \ \
  \vert \phi_{1} \rangle
  =&
  \frac{1}{2}
  (\vert 0\rangle_{i} \vert \! \uparrow\rangle_{j} \vert \! \downarrow\rangle_{k}
  -\vert \! \downarrow\rangle_{i} \vert \! \uparrow\rangle_{j} \vert 0\rangle_{k}
\nonumber \\
  &
  -\vert 0\rangle_{i} \vert \! \downarrow\rangle_{j} \vert \! \uparrow\rangle_{k}
  +\vert \! \uparrow\rangle_{i} \vert \! \downarrow\rangle_{j} \vert 0\rangle_{k})
\nonumber \\
  =&
  \frac{1}{\sqrt{2}}( \vert 0\rangle_{i} \vert s \rangle_{jk} + \vert s \rangle_{ij} \vert 0\rangle_{k}),
\end{align}
\begin{align}
  \lambda_{2} = 0,
  \ \
  \vert \phi_{2} \rangle
  =&
  c \vert 0\rangle_{i} \vert \! \uparrow\rangle_{j} \vert \! \downarrow\rangle_{k}
  +\sqrt{\frac{1}{2}-c^{2}} \vert \! \downarrow\rangle_{i} \vert \! \uparrow\rangle_{j} \vert 0\rangle_{k}
\nonumber \\
  &
  +c \vert 0\rangle_{i} \vert \! \downarrow\rangle_{j} \vert \! \uparrow\rangle_{k}
  +\sqrt{\frac{1}{2}-c^{2}} \vert \! \uparrow\rangle_{i} \vert \! \downarrow\rangle_{j} \vert 0\rangle_{k}
\nonumber \\
  =&
  \sqrt{2}c \vert 0\rangle_{i} \vert t \rangle_{jk} + \sqrt{1-2c^{2}} \vert t \rangle_{ij} \vert 0\rangle_{k},
\end{align}
\begin{align}
  \lambda_{3} = 0,
  \ \
  \vert \phi_{3} \rangle
  =&
  c \vert 0\rangle_{i} \vert \! \uparrow\rangle_{j} \vert \! \downarrow\rangle_{k}
  -\sqrt{\frac{1}{2}-c^{2}} \vert \! \downarrow\rangle_{i} \vert \! \uparrow\rangle_{j} \vert 0\rangle_{k}
\nonumber \\
  &
  +c \vert 0\rangle_{i} \vert \! \downarrow\rangle_{j} \vert \! \uparrow\rangle_{k}
  -\sqrt{\frac{1}{2}-c^{2}} \vert \! \uparrow\rangle_{i} \vert \! \downarrow\rangle_{j} \vert 0\rangle_{k}
\nonumber \\
  =&
  \sqrt{2}c \vert 0\rangle_{i} \vert t \rangle_{jk} - \sqrt{1-2c^{2}} \vert t \rangle_{ij} \vert 0\rangle_{k},
\end{align}
\begin{align}
  \lambda_{4} = \frac{J_{ijk}}{2},
  \ \
  \vert \phi_{4} \rangle
  =&
  \frac{1}{2}
  (\vert 0\rangle_{i} \vert \! \uparrow\rangle_{j} \vert \! \downarrow\rangle_{k}
  +\vert \! \downarrow\rangle_{i} \vert \! \uparrow\rangle_{j} \vert 0\rangle_{k}
\nonumber \\
  &
  -\vert 0\rangle_{i} \vert \! \downarrow\rangle_{j} \vert \! \uparrow\rangle_{k}
  -\vert \! \uparrow\rangle_{i} \vert \! \downarrow\rangle_{j} \vert 0\rangle_{k})
\nonumber \\
  =&
  \frac{1}{\sqrt{2}}( \vert 0\rangle_{i} \vert s \rangle_{jk} - \vert s \rangle_{ij} \vert 0\rangle_{k}),
\end{align}
where
$\vert s \rangle_{ij}$
and
$\vert t \rangle_{ij}$
denote, respectively, the spin singlet and triplet states on the bond $ij$,
\begin{align}
  \vert s \rangle_{ij} =&
  \frac{1}{\sqrt{2}}
  ( \vert \! \uparrow\rangle_{i} \vert \! \downarrow\rangle_{j}
  - \vert \! \downarrow\rangle_{i} \vert \! \uparrow\rangle_{j} ),
\\
  \vert t \rangle_{ij} =&
  \frac{1}{\sqrt{2}}
  ( \vert \! \uparrow\rangle_{i} \vert \! \downarrow\rangle_{j}
  + \vert \! \downarrow\rangle_{i} \vert \! \uparrow\rangle_{j} ),
\end{align}
and $c$ is a constant.

We remark that $\vert \phi_{1} \rangle$ and $\vert \phi_{4} \rangle$ are the superposition of
the spin singlet state on the bond $ij$ and the one on the bond $jk$,
while the sign is opposite.
Thus,
the lowest- and highest-energy states have the total spin zero.
Similarly,
$\vert \phi_{2} \rangle$ and $\vert \phi_{3} \rangle$ are the superposition of
the spin triplet state on the bond $ij$ and the one on the bond $jk$,
while the sign is opposite.
Thus, the two zero-energy states have the total spin one.


\end{document}